\documentclass[12pt,preprint]{aastex}

\newcommand{\asca}{{\sl ASCA}}
\newcommand{\exosat}{{\sl EXOSAT}}
\newcommand{\einstein}{{\sl Einstein}}
\newcommand{\rosat}{{\sl ROSAT}}
\newcommand{\rass}{{\sl RASS}}
\newcommand{\xmm}{XMM-{\sl{Newton}}}
\newcommand{\chandra}{{\sl Chandra}}
\newcommand{\scrA}{{a_P}}

\slugcomment{accepted for publication in ApJ}

\shorttitle{EGPs and X-rays}
\shortauthors{Kashyap et al.}

\begin{document}

\title{Extrasolar Giant Planets and X-ray Activity}

\author{Vinay L.\ Kashyap, Jeremy J.\ Drake, Steven H.\ Saar}
\affil{Harvard-Smithsonian Center for Astrophysics,
\\ 60 Garden Street, \\ Cambridge, MA 02138}
\email{vkashyap@cfa.harvard.edu \\
jdrake@cfa.harvard.edu \\
ssaar@cfa.harvard.edu}

\begin{abstract}

We have carried out a survey of X-ray emission from
stars with giant planets, combining both archival
and targeted surveys.
Over 230 stars have been currently identified as
possessing planets, and roughly a third of these
have been detected in X-rays.
We carry out detailed statistical analysis on
a volume limited sample of main sequence star
systems with detected planets, comparing
subsamples of stars that have close-in planets with
stars that have more distant planets.  This analysis
reveals strong
evidence that stars with close-in giant planets are
on average more X-ray active by a factor $\approx 4$
than those with planets that are more distant.
This result persists for various sample selections.
We find that even after accounting
for observational sample bias, a significant residual
difference still remains.  This observational
result is consistent with the hypothesis that giant
planets in close proximity to the primary stars
influences the stellar magnetic activity.

\end{abstract}

\keywords{stellar activity -- extrasolar planets -- magnetospheres -- surveys -- X-rays: general -- methods: statistical }

\section{Introduction}
\label{s:intro}

After centuries of ignorance on planetary systems beyond
our own, we are now in an era when stars are being routinely
identified as possessing planets.
Since the discovery by Mayor \& Queloz (1995) of a Jupiter-type
planet (mass $0.44~M_J$) orbiting close in to the G2 star 51\,Peg
($0.05$~AU), almost three hundred new Extrasolar Giant
Planets (EGPs) have been found.\footnote{
The current state of this rapidly advancing field is summarized
by the International Astronomical Union's Working Group on
Extrasolar Planets at {\tt http://www.ciw.edu/boss/IAU/div3/wgesp/}
and at the Exoplanet Encyclopedia at {\tt http://exoplanet.eu/}
}

One of the surprising results so far have been the detection of
planets with orbital radii, $\scrA$, much smaller than seen in
the Solar System, with values as low as $\scrA\approx$0.03~AU.  At
such small separations, it is likely that these giant planets will
have measurable tidal ($\propto \scrA^{-3}$) or magnetic ($\propto
\scrA^{-2}$ for large $\scrA$) effects on the primary stars (Cuntz, Saar,
\& Musielak 2000; Saar, Cuntz, Shkolnik 2004).  Both these effects increase non-linearly for
small values of $\scrA$, leading to potentially dramatic disruptions
of the stellar environment for stars with close-in giant planets.
How these disruptions affect coronal activity remains an open
question.  It is possible that activity will be enhanced because
of the magnetic interactions between star and planet.  Tidal
bulges will also have an effect on the stability and energetics of the chromosphere.
Hitherto, analyses of variables that control stellar activity
levels for single stars have focused on rotation rate, mass, age, evolutionary
state, and to some extent metallicity, as factors determining
coronal activity.  Now we are presented with another possibility,
viz., the existence of close-in EGPs may also be a controlling
parameter.  It is well known that binary stars with close stellar
companions are generally more active than single stars (see e.g.,
Pye et al.\ 1994).  Stars with close-in EGPs may be of the
same type, though it is unclear whether similar mechanisms of
activity enhancements hold at both extremes of this family.

A preliminary search for planet-induced stellar activity enhancement
was carried out by Bastian, Dulk \& Leblanc (2000) in the radio and
by Saar \& Cuntz (2001) in the optical.  But due to insufficient
sensitivity, neither study succeeded in uncovering clear evidence for
this phenomenon.  However, a sensitive search for activity
(Shkolnik et al.\ 2003; see also Cuntz \& Shkolnik 2002) detected
enhanced stellar emission in the Ca~II~HK lines from the chromosphere
of HD\,179949, in phase with the orbit of its close-in giant planet
($P_{\rm orb} = 3.09$ days, $M \sin i = 0.93 M_J$, $\scrA = 0.045$ AU).
The enhancement is clearly planet-related because it is not in phase
with the stellar rotation rate $P_{\rm rot} = 8 - 9$ days (Tinney et
al.\ 2001).  Because the Ca~II~HK enhancement  is seen only on the
hemisphere facing the planet, a magnetic interaction is preferred
over a tidal one.  Since a unipolar inductor model (viz., the Io--Jupiter
interaction; Zarka et al.\ 2001) does not fit the data as well (Saar
et al.\ 2004), this appears to be the first observational evidence
for an exoplanetary magnetosphere.  These observations also provide
an opportunity to probe the high-energy particle environment near
EGPs since the interaction strength depends on the magnitudes of the
stellar and planetary magnetic fields $B_*$ and $B_P$ (Saar et al.\
2004).  Note however that these emissions appear to be not phase
locked, as is expected if the stellar magnetic fields are varying
(McIvor, Jardine, \& Holzwarth 2006; Cranmer \& Saar 2006).

The peak strength of the HK emission flux enhancement found is
$\approx$4\% (Shkolnik et al.\ 2004).  Such low-level enhancements
are difficult to detect and study in any detail.  However, if the
emission scales like other manifestations of stellar activity,
coronal enhancements should be much larger: e.g., Ayres et al.\ (1995)
found surface fluxes $F_X/F_{bol} \propto (F_{\rm C\,IV}/F_{bol})^3$.
Follow-up observations of HD\,179949 in X-rays (\chandra/ACIS;
Saar et al.\ 2006) show significant spectral and temporal variability
phased to the planetary orbit, but there is some residual ambiguity
due to the poorly constrained stellar rotational period ($P_{\rm rot} 
\approx 7-10$ d) -- some fraction of the variation may be due to 
changes in the underlying stellar activity.

This picture has been reinforced by recent observations of similar
activity on $\upsilon$~And and a confirmation of the activity
synchronization on HD\,179949 (Shkolnik et al.\ 2008).  There are
also indications of the influence of a planet on stellar activity
in $\tau$~Boo and HD\,189733, though the results are inconclusive.
The newer observations also show that the possible influence
of the giant planet on stellar activity is complex, intermittent,
and prone to phase shifts (Saar et al.\ 2006, Shkolnik et al.\ 2008,
Lanza 2008).

In \S\ref{s:data}, we detail the X-ray data available for stars
with giant plants.  In \S\ref{s:analysis}, we carry out statistical
searches for trends in the data as a function of $\scrA$, and show
evidence of a significant deviation between extremal subsamples.
In these analyses, we take into account the large numbers of
censored data (that is, undetected stars) and properly include
their effect.
We discuss these results in \S\ref{s:discuss} with careful
attention to the biases in the sample.  We summarize our results
in \S\ref{s:summary}.

\section{Data}
\label{s:data}

To date,\footnote{
As of February 2008.
}
over 230 star systems have been identified as possessing planets;
various methods such as spectroscopically detecting the wobble due to planetary
orbits in radial velocity measurements (e.g., Butler et al.\ 1996)
and detecting photometric dips in the light curve due to disk transit
(e.g., Konacki et al.\ 2003,2004)
have been used in these identifications.
For the sake of simplicity and homogeneity (see \S\ref{s:bias}),
we shall use for our sample only those stars which have had planets
discovered or verified using the spectroscopic method.
These stars are listed in Table~\ref{t:egp_x}.
In this list, we do not include those stars in which low-mass stellar
or brown-dwarf companions have been detected using the above
methods.\footnote{
While such a study is intrinsically interesting on its own merits, it is
beyond the scope of our analysis.
Binary star systems tend to be more X-ray active
than single stars (e.g., Pye et al.\ 1994), possibly due to the
higher angular momentum available or due to the manner of the evolution
of the system, whereas the primary mechanisms for activity enhancements
for stars with close-in EGPs appears to be tidal or magnetic disruption.
Here we concentrate on a well-defined and limited sample of ostensibly
single stars to explore the possible changes in their properties.
}
We have searched for X-ray counterparts of these stars in archival
data from \asca, \exosat, \einstein, \rosat, \xmm, and \chandra\ missions.
Using the HEASARC Browse\footnote{
{\tt http://heasarc.gsfc.nasa.gov/db-perl/W3Browse/w3browse.pl}
}, we find matches in
the \asca\ Medium Sensitivity Survey ({\tt ascagis}; Ueda et al.\ 2001),
the \exosat\ results ({\tt sc\_cma\_view}; White \& Peacock 1988),
the \einstein\ 2-$\sigma$ catalog ({\tt twosigma}; Moran et al.\ 1996),
the \einstein/IPC source catalog ({\tt einstein2e}; Harris et al.\ 1994),
the \rosat/HRI complete results archive ({\tt roshritotal}; Voges et al.\ 2001),
the Brera Multi-scale Wavelet \rosat/HRI catalog ({\tt bmwhricat}; Panzera et al.\ 2003),
the \rosat/PSPC complete results archive ({\tt rospspctotal}; Voges et al.\ 2001),
the WGACAT ({\tt wgacat}; White, Giommi, \& Angelini 1994),
the \rosat\ All-Sky Survey (\rass) Bright Sources catalog ({\tt rassbsc}; Voges et al.\ 1996),
the \rass\ Faint Sources catalog ({\tt rassfsc}; Voges et al. 1999),
the \rass\ A-K Dwarfs/Subgiants catalog ({\tt rassdwarf}; H\"{u}nsch, Schmitt, \& Voges 1998),
and the XMM Serendipitous Source catalog ({\tt xmmssc}; Pye et al.\ 2006).
No matches have been found in the ChaMP database (Kim et al.\ 2004),
and with the exception of $\epsilon$\,Eri, none of these stars have
been observed with {\sl Chandra}.

{

\begin{deluxetable}{lccccccl}
\tabletypesize{\scriptsize}
\tablecolumns{9}
\tablewidth{500pt}
\tablecaption{Stars with Giant Planets \label{t:egp_x}}
\tablehead{
\colhead{Name} & \colhead{Spectral} & \colhead{$m_V$} & \colhead{$d$} & \colhead{$\scrA$\tablenotemark{a}} & \colhead{X-ray Flux\tablenotemark{b}} & \colhead{$\log_{10}{L_X}$} & \colhead{comments\tablenotemark{c}} \\
\colhead{} & \colhead{Type} & \colhead{} & \colhead{[pc]} & \colhead{[AU]} & \colhead{[$10^{-13}$~ergs~cm$^{-2}$~s$^{-1}$]} & [ergs~s$^{-1}$] & \colhead{}
}
\startdata
HD 41004 B & M2 &  6.67 &     43.0 & 0.018 & 11.6 $\pm$ 1.55 & $ 29.41 \pm   0.06$ & [138]  {\underline{P}}  \\
OGLE-TR-56 & G &  6.20 &     1500 & 0.023 & $<1.79$ & $< 30.26$ & [57] P  \\
TrES-3 & G &  5.52 &     292. & 0.023 & $<0.646$ & $< 29.82$ & [91] P  \\
OGLE-TR-113 & K & 20.15 &     1500 & 0.023 & $<1.80$ & $< 30.45$ & [58] P  \\
WASP-4 & G7V &  4.60 &     300. & 0.023 & $<0.646$ & $< 29.84$ & [135] P  \\
WASP-5 & G4V &  5.10 &     297. & 0.027 & $<0.646$ & $< 29.83$ & [1] P  \\
GJ 436 & M2.5 &  5.49 &     10.2 & 0.029 & 1.17 $\pm$ 0.525 & $ 27.16 \pm   0.19$ & [17]  {\underline{P}}  \\
SWEEPS-11 & F9 &  5.95 &     2000 & 0.030 & 64.9 $\pm$ 4.55 & $ 33.49 \pm   0.03$ & [102]  {\underline{P}}  \\
OGLE-TR-132 & F &  5.00 &     1500 & 0.031 & $<1.78$ & $< 30.45$ & [7] P  \\
WASP-2 & K1V &  9.16 &     207. & 0.031 & $<0.646$ & $< 29.52$ & [29] P  \\
HD 189733 & K1-K2 & 10.08 &     19.3 & 0.031 & 6.08 $\pm$ 0.183 & $ 28.43 \pm   0.01$ & [9]  {\underline{P}} E H X  \\
WASP-3 & F7V & 20.10 &     223. & 0.032 & $<0.646$ & $< 29.58$ & [96] P  \\
HD 212301 & F8 V & 13.60 &     52.7 & 0.036 & $<2.13$ & $< 28.85$ & [63] P  \\
HD 63454 & K4 V & 12.57 &     35.8 & 0.036 & 0.655 $\pm$ 0.285 & $ 28.00 \pm   0.19$ & [81]  {\underline{P}}  \\
TrES-2 & G0V &  3.73 &     220. & 0.037 & $<1.46$ & $< 29.93$ & [90] P  \\
XO-2 & K0V &  6.70 &     149. & 0.037 & $<0.646$ & $< 29.23$ & [10] P  \\
HD 73256 & G8/K0 & 13.00 &     36.5 & 0.037 & 2.15 $\pm$ 0.905 & $ 28.53 \pm   0.18$ & [129]  {\underline{P}}  \\
HAT-P-7 & F3 & 10.68 &     320. & 0.038 & $<0.646$ & $< 29.90$ & [93] P  \\
WASP-1 & F7V &  9.36 &     499. & 0.038 & $<0.646$ & $< 30.28$ & [29] P  \\
HAT-P-3 & K & 11.40 &     140. & 0.039 & $<0.646$ & $< 29.18$ & [124] P  \\
HD 86081 & F8V &  3.23 &     91.0 & 0.039 & $<1.46$ & $< 29.16$ & [50] P  \\
GJ 674 & M2.5 & 10.42 &     4.54 & 0.039 & 13.9 $\pm$ 2.05 & $ 27.53 \pm   0.06$ & [68]  {\underline{P}} N  \\
TrES-1 & K0V & 10.56 &     157. & 0.039 & $<1.14$ & $< 29.53$ & [105] P  \\
HD 83443 & K0 V &  6.17 &     43.5 & 0.041 & $<1.70$ & $< 28.59$ & [76] P  \\
HAT-P-5 & K & 10.17 &     340. & 0.041 & $<0.646$ & $< 29.95$ & [3] P  \\
Gl 581 & M3 & 10.40 &     6.26 & 0.041/0.073/0.25 & $<1.64$ & $< 26.89$ & [6] P  \\
HD 46375 & K1 IV &  8.71 &     33.4 & 0.041 & 0.212 $\pm$ 0.036 & $ 27.45 \pm   0.07$ & [72]  {\underline{N}}  \\
TW Hya & K8V & 11.86 &     54.0 & 0.041 & 42.7 $\pm$ 0.147 & $ 30.17 \pm   0.00$ & [115]  {\underline{N}} P A  \\
OGLE-TR-10 & G or K & 11.20 &     1500 & 0.042 & 0.081 $\pm$ 0.047 & $ 30.34 \pm   0.25$ & [8]  {\underline{N}}  \\
HD 187123 & G5 & 12.00 &     50.0 & 0.042 & $<0.073$ & $< 27.34$ & [13] N  \\
HD 330075 & G5 & 10.50 &     50.2 & 0.043 & 0.047 $\pm$ 0.032 & $ 27.15 \pm   0.30$ & [75]  {\underline{N}}  \\
HD 149026 & G0 IV & 10.50 &     78.9 & 0.043 & $<1.15$ & $< 28.93$ & [108] P  \\
HD 2638 & G5 &  8.42 &     53.7 & 0.044 & $<0.987$ & $< 28.53$ & [81] P  \\
HAT-P-4 & F &  8.21 &     310. & 0.045 & $<0.646$ & $< 29.87$ & [60] P  \\
HD 209458 & G0 V &  7.47 &     47.0 & 0.045 & 0.039 $\pm$ 0.018 & $ 27.02 \pm   0.20$ & [22]  {\underline{N}}  \\
HD 179949 & F8 V &  8.06 &     27.0 & 0.045 & 4.68 $\pm$ 1.12 & $ 28.61 \pm   0.10$ & [120]  {\underline{P}}  \\
$\tau$ Boo & F7 V &  5.80 &     15.0 & 0.046 & 25.1 $\pm$ 0.0886 & $ 28.83 \pm   0.00$ & [12]  {\underline{N}} P E H A  \\
HD 75289 & G0 V &  7.36 &     28.9 & 0.046 & $<1.68$ & $< 28.23$ & [125] P  \\
BD-10 3166 & G4 V &  5.52 &     100. & 0.046 & 3.14 $\pm$ 1.01 & $ 29.58 \pm   0.14$ & [14]  {\underline{P}}  \\
Lupus-TR-3 & K1V &  6.29 &     1780 & 0.046 & $<0.646$ & $< 30.07$ & [134] P  \\
OGLE-TR-111 & G8 &  8.02 &     1500 & 0.047 & $<0.646$ & $< 29.69$ & [105] P  \\
HD 88133 & G5 IV &  7.00 &     74.5 & 0.047 & $<1.08$ & $< 28.86$ & [40] P  \\
XO-3 & F5V &  8.74 &     260. & 0.048 & $<0.646$ & $< 29.72$ & [49] P  \\
XO-1 & G1V &  8.20 &     200. & 0.049 & $<0.646$ & $< 29.49$ & [78] P  \\
TrES-4 & F &  7.61 &     440. & 0.049 & $<1.50$ & $< 30.54$ & [67] P  \\
HD 76700 & G6 V &  8.80 &     59.7 & 0.049 & $<2.20$ & $< 28.97$ & [122] P  \\
HD 102195 & K0V &  6.69 &     29.0 & 0.049 & 1.06 $\pm$ 0.423 & $ 28.03 \pm   0.17$ & [45]  {\underline{P}}  \\
OGLE-TR-182 & G2 &  7.30 &     2618 & 0.051 & $<1.76$ & $< 30.63$ & [97] P  \\
OGLE-TR-211 & F4 &  7.57 &     2177 & 0.051 & $<1.83$ & $< 31.28$ & [117] P  \\
HD 219828 & G0IV &  7.54 &     81.1 & 0.052 & $<0.989$ & $< 28.89$ & [79] P  \\
51 Peg & G2 IV &  7.30 &     14.7 & 0.052 & 0.247 $\pm$ 0.052 & $ 26.80 \pm   0.09$ & [74]  {\underline{P}}  \\
HAT-P-6 & F &  7.18 &     200. & 0.052 & $<0.646$ & $< 29.49$ & [89] P  \\
HD 149143 & G0 IV &  8.07 &     63.0 & 0.053 & $<1.47$ & $< 28.84$ & [39] P  \\
SWEEPS-04 & F5 &  6.42 &     2000 & 0.055 & $<0.646$ & $< 29.61$ & [102] P  \\
HAT-P-1 & G0V &  4.70 &     139. & 0.055 & $<0.646$ & $< 29.17$ & [68] P  \\
HD 49674 & G5 V &  7.60 &     40.7 & 0.058 & $<1.59$ & $< 28.50$ & [16] P  \\
$\upsilon$ And & F8 V &  5.48 &     13.5 & 0.059/0.83/2.5 & 5.98 $\pm$ 0.920 & $ 28.11 \pm   0.07$ & [12]  {\underline{P}}  \\
HD 109749 & G3 IV &  8.33 &     59.0 & 0.064 & $<1.73$ & $< 28.86$ & [39] P  \\
HD 168746 & G5 &  7.44 &     43.1 & 0.065 & $<1.55$ & $< 28.54$ & [94] P  \\
HAT-P-2 & F8 &  7.51 &     135. & 0.068 & $<0.646$ & $< 29.15$ & [68] P  \\
HIP 14810 & G5 &  8.05 &     52.9 & 0.069/0.41 & $<1.82$ & $< 28.78$ & [18] P  \\
HD 118203 & K0 &  7.57 &     88.6 & 0.070 & $<1.81$ & $< 29.23$ & [139] P  \\
HD 68988 & G0 &  8.45 &     58.0 & 0.071 & $<1.77$ & $< 28.85$ & [132] P  \\
HD 162020 & K2 V &  6.45 &     31.3 & 0.072 & 12.3 $\pm$ 2.58 & $ 29.16 \pm   0.09$ & [127]  {\underline{P}}  \\
HD 285968 & M2.5V &  6.94 &     9.40 & 0.073 & 2.87 $\pm$ 0.889 & $ 27.48 \pm   0.13$ & [33]  {\underline{P}}  \\
HD 217107 & G8 IV &  7.25 &     37.0 & 0.073/4.4 & 0.203 $\pm$ 0.075 & $ 27.52 \pm   0.16$ & [34]  {\underline{N}}  \\
HD 185269 & G0IV &  5.70 &     47.0 & 0.077 & $<2.17$ & $< 28.76$ & [51] P  \\
HD 69830 & K0V &  7.69 &     12.6 & 0.079/0.19/0.63 & 1.56 $\pm$ 0.535 & $ 27.47 \pm   0.15$ & [66]  {\underline{P}}  \\
HD 130322 & K0 V &  7.34 &     30.0 & 0.088 & 0.361 $\pm$ 0.050 & $ 27.59 \pm   0.06$ & [125]  {\underline{N}}  \\
HD 108147 & F8/G0 V &  5.38 &     38.6 & 0.104 & 0.356 $\pm$ 0.167 & $ 27.80 \pm   0.20$ & [94]  {\underline{N}}  \\
Gl 86 & K1V &  8.16 &     11.0 & 0.110 & $<0.646$ & $< 26.97$ & [98] P  \\
HD 4308 & G5 V &  7.90 &     21.9 & 0.114 & $<2.06$ & $< 28.07$ & [130] P  \\
55 Cnc & G8 V &  7.03 &     13.4 & 0.115/0.24/5.8/0.038 & 0.493 $\pm$ 0.126 & $ 27.02 \pm   0.11$ & [12]  {\underline{E}}  \\
HD 27894 & K2 V &  6.74 &     42.4 & 0.122 & $<4.51$ & $< 28.99$ & [81] P  \\
HD 99492 & K2V &  7.25 &     18.0 & 0.123 & 0.937 $\pm$ 0.374 & $ 27.56 \pm   0.17$ & [71]  {\underline{P}}  \\
HD 38529 & G4 IV &  7.27 &     42.4 & 0.129/3.7 & 4.96 $\pm$ 0.709 & $ 29.03 \pm   0.06$ & [35]  {\underline{P}}  \\
HD 195019 & G3 IV-V &  6.51 &     20.0 & 0.139 & 0.052 $\pm$ 0.020 & $ 26.40 \pm   0.17$ & [34]  {\underline{N}}  \\
HD 192263 & K2 V &  7.24 &     19.9 & 0.150 & 2.26 $\pm$ 1.06 & $ 28.03 \pm   0.20$ & [103]  {\underline{P}}  \\
HD 6434 & G3 IV &  5.15 &     40.3 & 0.150 & $<3.10$ & $< 28.78$ & [76] P  \\
HD 102117 & G6V &  6.78 &     42.0 & 0.153 & $<1.98$ & $< 28.62$ & [65] P  \\
HD 17156 & G0 &  7.28 &     78.2 & 0.159 & $<2.06$ & $< 29.18$ & [46] P  \\
HD 33283 & G3V &  9.18 &     86.0 & 0.168 & $<1.99$ & $< 29.24$ & [50] P  \\
Gliese 876 & M4 V &  6.99 &     4.72 & 0.208/0.13/0.021 & 0.837 $\pm$ 0.165 & $ 26.35 \pm   0.09$ & [30]  {\underline{P}} X  \\
$\rho$ CrB & G0V or G2V &  5.95 &     17.4 & 0.220 & $<1.36$ & $< 27.69$ & [12] P  \\
HD 11964 & G5 &  6.92 &     34.0 & 0.229/3.2 & $<1.53$ & $< 28.32$ & [18] P  \\
HD 224693 & G2IV &  7.95 &     94.0 & 0.233 & $<2.41$ & $< 29.41$ & [50] P  \\
HD 43691 & G0IVV = 8.03 &  5.91 &     93.2 & 0.240 & $<1.58$ & $< 29.21$ & [46] P  \\
HD 37605 & K0V &  8.22 &     42.9 & 0.250 & $<2.67$ & $< 28.77$ & [27] P  \\
HD 107148 & G5 &  7.74 &     51.3 & 0.269 & $<0.547$ & $< 28.24$ & [18] P  \\
HD 117618 & G2V &  8.31 &     38.0 & 0.280 & $<1.33$ & $< 28.36$ & [123] P  \\
HD 3651 & K0 V &  8.17 &     11.0 & 0.284 & 1.23 $\pm$ 0.232 & $ 27.25 \pm   0.08$ & [38]  {\underline{H}}  \\
HD 74156 & G0 &  8.03 &     64.6 & 0.294/3.8/1.0 & $<1.30$ & $< 28.81$ & [84] P  \\
HD 219449 & K0 III &  7.18 &     45.0 & 0.300 & $<0.619$ & $< 28.18$ & [80] P  \\
HD 114762 & F9V &  7.98 &     39.5 & 0.300 & 0.376 $\pm$ 0.049 & $ 27.85 \pm   0.06$ & [62]  {\underline{N}}  \\
HD 168443 & G5 &  6.25 &     37.9 & 0.300/2.9 & $<1.56$ & $< 28.43$ & [127] P  \\
HD 101930 & K1 V &  7.86 &     30.5 & 0.302 & $<1.94$ & $< 28.33$ & [65] P  \\
HD 121504 & G2 V &  6.68 &     44.4 & 0.320 & $<1.41$ & $< 28.52$ & [76] P  \\
HD 178911 B & G5 &  7.22 &     46.7 & 0.320 & 1.66 $\pm$ 0.633 & $ 28.64 \pm   0.17$ & [137]  {\underline{P}}  \\
HD 16141 & G5 IV &  7.83 &     35.9 & 0.350 & $<2.46$ & $< 28.58$ & [72] P  \\
HD 80606 & G5 &  8.22 &     58.4 & 0.439 & $<1.80$ & $< 28.86$ & [82] P  \\
HD 216770 & K1 V &  7.68 &     38.0 & 0.460 & $<1.96$ & $< 28.53$ & [76] P  \\
HD 93083 & K3 V &  7.31 &     28.9 & 0.477 & $<1.34$ & $< 28.13$ & [65] P  \\
70 Vir & G4 V &  5.71 &     22.0 & 0.480 & 0.458 $\pm$ 0.134 & $ 27.42 \pm   0.13$ & [69]  {\underline{H}}  \\
GJ 3021 & G6 V &  7.70 &     17.6 & 0.490 & 28.3 $\pm$ 3.63 & $ 29.02 \pm   0.06$ & [83]  {\underline{P}} N  \\
HD 52265 & G0 V &  7.79 &     28.0 & 0.490 & $<2.05$ & $< 28.28$ & [14] P  \\
HD 208487 & G2V &  6.45 &     45.0 & 0.490 & $<4.35$ & $< 29.02$ & [123] P  \\
HD 37124 & G4 V &  6.91 &     33.0 & 0.530/3.2/1.6 & $<1.70$ & $< 28.34$ & [131] P  \\
HD 231701 & F8V &  7.60 &     108. & 0.556 & $<2.10$ & $< 29.47$ & [41] P  \\
HD 155358 & G0 &  6.40 &     42.7 & 0.628/1.2 & $<1.69$ & $< 28.57$ & [28] P  \\
HD 73526 & G6 V &  5.06 &     99.0 & 0.660/1.0 & $<1.56$ & $< 29.26$ & [122] P  \\
ksi Aql & G9IIIb &  8.08 &     62.7 & 0.680 & $<123.$ & $< 30.76$ & [110] P  \\
HD 75898 & G0 &  6.41 &     80.6 & 0.737 & $<1.35$ & $< 29.02$ & [101] P  \\
HD 8574 & F8 &  9.01 &     44.2 & 0.760 & $<3.75$ & $< 28.94$ & [95] P  \\
HD 134987 & G5 V &  7.38 &     25.0 & 0.780 & $<1.31$ & $< 27.99$ & [131] P  \\
HD 104985 & G9 III &  7.48 &     102. & 0.780 & $<1.47$ & $< 29.26$ & [107] P  \\
HD 81688 & K0III-IV &  7.65 &     88.3 & 0.810 & $<2.74$ & $< 29.41$ & [110] P  \\
HD 169830 & F8 V &  6.63 &     36.3 & 0.810/3.6 & 0.887 $\pm$ 0.053 & $ 28.15 \pm   0.03$ & [76]  {\underline{P}}  \\
HD 40979 & F8 V &  5.94 &     33.3 & 0.811 & $<1.64$ & $< 28.34$ & [36] P  \\
HD 150706 & G0 &  7.77 &     27.2 & 0.820 & 8.57 $\pm$ 0.714 & $ 28.88 \pm   0.04$ & [126]  {\underline{P}} H  \\
HD 202206 & G6 V &  6.80 &     46.3 & 0.830/2.5 & $<1.61$ & $< 28.62$ & [127] P  \\
HD 12661 & G6 V &  6.03 &     37.2 & 0.830/2.6 & $<1.10$ & $< 28.26$ & [35] P  \\
4 Uma & K1III &  6.06 &     62.4 & 0.870 & 1.47 $\pm$ 0.101 & $ 28.84 \pm   0.03$ & [31]  {\underline{H}} P  \\
HD 89744 & F7 V &  8.10 &     40.0 & 0.890 & 0.597 $\pm$ 0.079 & $ 28.06 \pm   0.06$ & [59]  {\underline{P}}  \\
HR 810 & G0V pecul. &  6.18 &     15.5 & 0.910 & 19.1 $\pm$ 2.73 & $ 28.74 \pm   0.06$ & [61]  {\underline{P}}  \\
HD 59686 & K2 III &  4.21 &     92.0 & 0.911 & $<1.86$ & $< 29.28$ & [80] P  \\
GJ 317 & M3.5 &  8.04 &     9.17 & 0.950 & $<3.20$ & $< 27.25$ & [52] P  \\
HD 92788 & G5 &  7.83 &     32.8 & 0.970 & $<0.963$ & $< 28.09$ & [35] P  \\
HD 142 & G1 IV &  7.70 &     20.6 & 0.980 & $<3.13$ & $< 28.20$ & [121] P  \\
HD 156846 & G0V &  8.23 &     49.0 & 0.990 & $<1.91$ & $< 28.74$ & [119] P  \\
HD 177830 & K0 &  7.10 &     59.0 & 1.000 & $<1.77$ & $< 28.87$ & [131] P  \\
ChaHa8 & M6.5 &  8.58 &     160. & 1.000 & 53.7 $\pm$ 2.02 & $ 31.22 \pm   0.02$ & [48]  {\underline{A}} N P  \\
HD 122430 & K3III &  8.99 &     135. & 1.020 & $<1.61$ & $< 29.54$ & [111] P  \\
HD 28185 & G5 &  7.24 &     39.4 & 1.030 & $<2.21$ & $< 28.61$ & [104] P  \\
HD 175541 & G8IV &  9.80 &     128. & 1.030 & $<2.95$ & $< 29.76$ & [46] P  \\
HD 100777 & K0 &  4.44 &     52.8 & 1.030 & $<0.504$ & $< 28.23$ & [86] P  \\
HD 142415 & G1 V &  9.42 &     34.2 & 1.050 & 2.88 $\pm$ 0.577 & $ 28.61 \pm   0.09$ & [76]  {\underline{P}}  \\
HD 108874 & G5 &  7.81 &     68.5 & 1.051/2.7 & $<0.603$ & $< 28.53$ & [16] P  \\
HD 4203 & G5 &  9.97 &     77.5 & 1.090 & 0.629 $\pm$ 0.294 & $ 28.66 \pm   0.20$ & [132]  {\underline{E}}  \\
HD 128311 & K0 &  8.41 &     16.6 & 1.099/1.8 & 10.2 $\pm$ 1.35 & $ 28.52 \pm   0.06$ & [16]  {\underline{P}} H U  \\
HD 33564 & F6 V &  9.00 &     21.0 & 1.100 & 1.30 $\pm$ 0.455 & $ 27.84 \pm   0.15$ & [43]  {\underline{P}}  \\
HD 210277 & G0 &  8.05 &     21.3 & 1.100 & $<1.32$ & $< 27.85$ & [70] P  \\
HD 99109 & K0 &  5.10 &     60.5 & 1.105 & $<0.548$ & $< 28.38$ & [18] P  \\
HD 192699 & G8IV &  5.80 &     67.0 & 1.160 & $<4.36$ & $< 29.37$ & [68] P  \\
HD 210702 & K1III &  7.68 &     56.0 & 1.170 & $<1.45$ & $< 28.73$ & [68] P  \\
HD 27442 & K2 IV a &  8.69 &     18.1 & 1.180 & 0.826 $\pm$ 0.304 & $ 27.51 \pm   0.16$ & [15]  {\underline{P}}  \\
HD 82943 & G0 &  5.94 &     27.5 & 1.190/0.75 & $<1.13$ & $< 28.01$ & [76] P  \\
HD 188015 & G5IV &  5.67 &     52.6 & 1.190 & $<2.46$ & $< 28.91$ & [71] P  \\
HD 125612 & G3V &  6.73 &     52.8 & 1.200 & $<1.48$ & $< 28.69$ & [41] P  \\
HD 114783 & K0 &  8.65 &     22.0 & 1.200 & $<0.861$ & $< 27.70$ & [132] P  \\
HD 154857 & G5V &  8.65 &     68.5 & 1.200 & $<2.20$ & $< 29.09$ & [77] P  \\
HD 221287 & F7 V &  7.91 &     52.9 & 1.250 & 3.35 $\pm$ 1.25 & $ 29.05 \pm   0.16$ & [68]  {\underline{P}}  \\
HD 20367 & G0 &  8.69 &     27.0 & 1.250 & 20.8 $\pm$ 0.165 & $ 29.26 \pm   0.00$ & [126]  {\underline{N}} P  \\
HD 147513 & G3/G5V &  7.79 &     12.9 & 1.260 & 51.5 $\pm$ 1.16 & $ 29.01 \pm   0.01$ & [76]  {\underline{P}} X  \\
HIP 75458 & K2 III &  6.54 &     31.5 & 1.275 & 0.228 $\pm$ 0.065 & $ 27.43 \pm   0.12$ & [42]  {\underline{P}}  \\
HD 4113 & G5V &  8.03 &     44.0 & 1.280 & $<5.14$ & $< 29.08$ & [119] P  \\
HD 171028 & G0 &  7.88 &     90.0 & 1.290 & $<2.63$ & $< 29.41$ & [106] P  \\
HD 17092 & K0III &  7.84 &     109. & 1.290 & $<1.64$ & $< 29.37$ & [88] P  \\
HD 19994 & F8 V &  5.26 &     22.4 & 1.300 & $<2.25$ & $< 28.13$ & [76] P  \\
HD 167042 & K1III &  8.10 &     50.0 & 1.300 & $<1.73$ & $< 28.71$ & [46] P  \\
HD 41004 A & K1 V &  7.22 &     42.5 & 1.310 & 11.6 $\pm$ 1.55 & $ 29.40 \pm   0.06$ & [128]  {\underline{P}}  \\
HD 222582 & G5 &  6.86 &     42.0 & 1.350 & $<0.773$ & $< 28.21$ & [131] P  \\
HD 20782 & G2 V &  6.30 &     36.0 & 1.360 & $<1.68$ & $< 28.42$ & [56] P  \\
HD 65216 & G5 V &  8.05 &     34.3 & 1.370 & $<0.646$ & $< 27.96$ & [76] P  \\
HD 160691 & G3 IV-V &  5.45 &     15.3 & 1.500/4.2/0.090/0.92 & 0.423 $\pm$ 0.061 & $ 27.07 \pm   0.06$ & [15]  {\underline{P}}  \\
HD 141937 & G2/G3 V &  1.15 &     33.5 & 1.520 & $<1.38$ & $< 28.27$ & [127] P  \\
HD 183263 & G2IV &  9.40 &     53.0 & 1.520 & $<3.71$ & $< 29.10$ & [71] P  \\
HD 47536 & K1 III &  7.72 &     121. & 1.610/5.2 & $<1.83$ & $< 29.51$ & [112] P  \\
HD 114386 & K3 V &  7.98 &     28.0 & 1.620 & $<1.12$ & $< 28.02$ & [76] P  \\
HD 23079 & F8/G0 V &  8.00 &     34.8 & 1.650 & $<3.86$ & $< 28.75$ & [121] P  \\
HD 4208 & G5 V &  8.21 &     33.9 & 1.670 & $<1.51$ & $< 28.32$ & [132] P  \\
16 Cyg B & G2.5 V &  5.95 &     21.4 & 1.680 & $<1.46$ & $< 27.90$ & [25] P  \\
HD 62509 & K0IIIb &  8.70 &     10.3 & 1.690 & 0.687 $\pm$ 0.022 & $ 26.94 \pm   0.01$ & [100]  {\underline{N}} P E A  \\
V391 Peg & sdB &  7.18 &     1400 & 1.700 & $<1.55$ & $< 31.56$ & [116\tablenotemark{d}] P  \\
HD 5319 & G5IV &  7.48 &     100. & 1.750 & $<1.60$ & $< 29.28$ & [101] P  \\
HD 70573 & G1-1.5V &  8.08 &     45.7 & 1.760 & 2.57 $\pm$ 0.423 & $ 28.81 \pm   0.07$ & [114]  {\underline{E}} P  \\
OGLE-05-071L & M5 &  9.00 &     2900 & 1.800 & 1.05 $\pm$ 0.189 & $ 32.03 \pm   0.08$ & [68\tablenotemark{d}]  {\underline{P}}  \\
HD 13189 & K2 II &  7.61 &     185. & 1.850 & $<1.59$ & $< 29.81$ & [47] P  \\
HD 45350 & G5 IV &  6.36 &     49.0 & 1.920 & $<1.60$ & $< 28.66$ & [71] P  \\
eps Tau & K0 III &  8.04 &     45.0 & 1.930 & 0.615 $\pm$ 0.175 & $ 28.17 \pm   0.12$ & [109]  {\underline{P}}  \\
HD 11977 & G8.5 III &  8.13 &     66.5 & 1.930 & $<2.83$ & $< 29.17$ & [113] P  \\
HD 81040 & G2/G3 &  8.93 &     32.6 & 1.940 & 1.21 $\pm$ 0.508 & $ 28.19 \pm   0.18$ & [118]  {\underline{P}}  \\
HD 111232 & G8V &  7.72 &     29.0 & 1.970 & $<2.18$ & $< 28.34$ & [76] P  \\
HD 132406 & G0V &  5.41 &     71.0 & 1.980 & $<1.91$ & $< 29.06$ & [46] P  \\
HD 159868 & G5V &  6.54 &     52.7 & 2.000 & $<2.11$ & $< 28.84$ & [92] P  \\
HD 213240 & G4 IV &  8.24 &     40.8 & 2.030 & $<1.84$ & $< 28.56$ & [104] P  \\
Gamma Cephei & K2 V &  7.11 &     11.8 & 2.044 & 0.550 $\pm$ 0.068 & $ 26.96 \pm   0.05$ & [26]  {\underline{P}}  \\
HD 187085 & G0 V &  8.74 &     45.0 & 2.050 & $<2.14$ & $< 28.71$ & [56] P  \\
HD 16175 & G0 &  8.06 &     59.8 & 2.070 & $<1.58$ & $< 28.83$ & [46] P  \\
HD 190647 & G5 &  7.01 &     54.2 & 2.070 & $<2.01$ & $< 28.85$ & [68] P  \\
HD 114729 & G3 V &  5.74 &     35.0 & 2.080 & $<1.24$ & $< 28.26$ & [16] P  \\
NGC 2423 3 & KIII &  7.10 &     766. & 2.100 & $<2.89$ & $< 31.31$ & [64] P  \\
OGLE-05-390L & M4 &  8.33 &     6500 & 2.100 & $<0.646$ & $< 30.82$ & [68\tablenotemark{d}] P  \\
HD 10647 & F8V &  8.80 &     17.3 & 2.100 & 4.51 $\pm$ 0.260 & $ 28.21 \pm   0.02$ & [128]  {\underline{P}}  \\
HD 164922 & K0V &  7.57 &     21.9 & 2.110 & $<1.54$ & $< 27.95$ & [18] P  \\
47 Uma & G0V &  8.51 &     14.0 & 2.110/7.7 & 0.583 $\pm$ 0.050 & $ 27.13 \pm   0.04$ & [11]  {\underline{N}}  \\
HD 10697 & G5 IV &  3.31 &     32.6 & 2.130 & 1.30 $\pm$ 0.520 & $ 28.22 \pm   0.17$ & [131]  {\underline{P}}  \\
HD 2039 & G2/G3 IV-V &  5.40 &     89.8 & 2.190 & $<2.68$ & $< 29.41$ & [122] P  \\
HD 170469 & G5IV & 17.40 &     65.0 & 2.240 & $<677.$ & $< 30.87$ & [41] P  \\
OGLE-06-109L & K9 & 10.04 &     1490 & 2.300/4.6 & $<0.646$ & $< 28.81$ & [44\tablenotemark{d}] P  \\
HD 136118 & F9 V &  7.40 &     52.3 & 2.300 & $<1.71$ & $< 28.75$ & [37] P  \\
HD 190228 & G5IV & 23.00 &     66.1 & 2.310 & $<2.52$ & $< 29.12$ & [95] P  \\
HD 11506 & G0V & 22.30 &     53.8 & 2.350 & $<1.70$ & $< 28.77$ & [41] P  \\
Gj 849 & M3.5 & 18.34 &     8.80 & 2.350 & 1.91 $\pm$ 0.486 & $ 27.25 \pm   0.11$ & [19]  {\underline{E}}  \\
HD 50554 & F8 & 20.17 &     31.0 & 2.380 & $<0.076$ & $< 26.94$ & [95] N  \\
NGC 4349 No 127 & KIII & 15.78 &     2176 & 2.380 & $<0.646$ & $< 31.56$ & [64] P  \\
HD 23127 & G2V & 17.98 &     89.1 & 2.400 & $<4.88$ & $< 29.67$ & [92] P  \\
HD 196050 & G3 V & 16.08 &     46.9 & 2.500 & $<2.25$ & $< 28.77$ & [76] P  \\
18 Del & G6III & 16.08 &     73.1 & 2.600 & 4.18 $\pm$ 0.836 & $ 29.43 \pm   0.09$ & [110]  {\underline{P}}  \\
HD 106252 & G0 & 16.84 &     37.4 & 2.610 & $<0.287$ & $< 27.68$ & [95] P  \\
HD 196885 & F8V & 14.80 &     33.0 & 2.630 & $<1.83$ & $< 28.38$ & [18] P  \\
HD 216435 & G0 V & 16.56 &     33.3 & 2.700 & 1.08 $\pm$ 0.290 & $ 28.16 \pm   0.12$ & [55]  {\underline{N}}  \\
kappa CrB & K1IVa & 21.20 &     31.1 & 2.700 & $<1.29$ & $< 28.17$ & [53] P  \\
HD 216437 & G4 IV-V & 99.00 &     26.5 & 2.700 & 0.179 $\pm$ 0.055 & $ 27.18 \pm   0.13$ & [76]  {\underline{N}}  \\
HD 23596 & F8 & 21.30 &     52.0 & 2.720 & $<1.59$ & $< 28.71$ & [95] P  \\
14 Her & K0 V & 17.40 &     18.1 & 2.770 & $<0.528$ & $< 27.32$ & [73] N  \\
OGLE-05-169L & M0 & 18.80 &     2700 & 2.800 & $<0.646$ & $< 28.47$ & [68\tablenotemark{d}] P  \\
HD 142022 A & K0 V & 19.83 &     35.9 & 2.800 & $<1.93$ & $< 28.47$ & [32] P  \\
HD 66428 & G5 & 11.10 &     55.0 & 3.180 & $<1.90$ & $< 28.84$ & [18] P  \\
HD 39091 & G1 IV &  4.50 &     20.5 & 3.290 & 0.423 $\pm$ 0.050 & $ 27.33 \pm   0.05$ & [54]  {\underline{P}}  \\
HD 70642 & G5 IV-V &  4.50 &     29.0 & 3.300 & 0.058 $\pm$ 0.019 & $ 26.77 \pm   0.14$ & [21]  {\underline{N}}  \\
$\epsilon$ Eri & K2 V &  4.50 &     3.20 & 3.390 & 108. $\pm$ 0.360 & $ 28.12 \pm   0.00$ & [20]  {\underline{N}} P E H X A U  \\
HD 117207 & G8VI/V & 12.40 &     33.0 & 3.780 & $<1.21$ & $< 28.20$ & [71] P  \\
HD 30177 & G8 V &  4.50 &     55.0 & 3.860 & $<3.20$ & $< 29.06$ & [122] P  \\
HD 50499 & G IV &  4.50 &     47.3 & 3.860 & $<3.70$ & $< 28.99$ & [133] P  \\
HD 190360 & G6 IV &  4.09 &     15.9 & 3.920/0.13 & $<0.286$ & $< 26.94$ & [85] N  \\
HD 89307 & G0V &  4.09 &     33.0 & 4.150 & $<0.883$ & $< 28.06$ & [18] P  \\
HD 72659 & G0 V & 11.79 &     51.4 & 4.160 & $<1.92$ & $< 28.78$ & [16] P  \\
HD 154345 & G8V & 11.98 &     18.1 & 4.190 & $<1.49$ & $< 27.76$ & [68] P  \\
SCR 1845 & M8.5 V & 10.64 &     3.85 & 4.500 & 3.20 $\pm$ 0.254 & $ 26.75 \pm   0.03$ & [5\tablenotemark{d}]  {\underline{P}}  \\
OGLE235-MOA53 & K5 & 12.60 &     6025 & 5.100 & $<0.646$ & $< 29.61$ & [68\tablenotemark{d}] P  \\
2M1207 & M8 & 12.26 &     52.4 & 46.000 & $<0.646$ & $< 25.91$ & [23\tablenotemark{d}] P  \\
GQ Lup & K7eV & 11.30 &     140. & 103.00 & 2.53 $\pm$ 0.397 & $ 29.77 \pm   0.07$ & [87\tablenotemark{d}]  {\underline{P}}  \\
AB Pic & K2 V & 11.18 &     45.6 & 275.00 & 39.8 $\pm$ 1.69 & $ 30.00 \pm   0.02$ & [24\tablenotemark{d}]  {\underline{P}} N E U  \\
UScoCTIO 108 & M7 &  9.80 &     145. & 670.00 & $<1.40$ & $< 29.55$ & [2\tablenotemark{d}] P  \\
\enddata
\tablenotetext{a}{Semi-major axes of planetary orbits}
\tablenotetext{b}{The X-ray flux adopted from the best available measurement (see text)}
\tablenotetext{c}{The reference for the planetary detection (see below), and the X-ray mission in which the star was detected: A={\sl ASCA}/GIS, X={\sl EXOSAT}/LE, H={\sl ROSAT}/HRI, E={\sl Einstein}/IPC, P={\sl ROSAT}/PSPC, N={\sl XMM-Newton}}
\tablenotetext{d}{Planets around stars have not been detected via the Radial Velocity method}
\tablerefs{
[1] Anderson et al.\ (2008);
[2] B\'{e}jar et al.\ (2008);
[3] Bakos et al.\ (2007);
[4] Barge et al.\ (2007);
[5] Biller et al.\ (2006);
[6] Bonfils et al.\ (2005);
[7] Bouchy et al.\ (2004);
[8] Bouchy et al.\ (2005a);
[9] Bouchy et al.\ (2005b);
[10] Burke et al.\ (2007);
[11] Butler \& Marcy (1996);
[12] Butler et al.\ (1997);
[13] Butler et al.\ (1998);
[14] Butler et al.\ (2000);
[15] Butler et al.\ (2001);
[16] Butler et al.\ (2003);
[17] Butler et al.\ (2004);
[18] Butler et al.\ (2006a);
[19] Butler et al.\ (2006b);
[20] Campbell, Walker, \& Yang (1998);
[21] Carter et al.\ (2003);
[22] Charbonneau et al.\ (2000);
[23] Chauvin et al.\ (2005a);
[24] Chauvin et al.\ (2005b);
[25] Cochran et al.\ (1997);
[26] Cochran et al.\ (2002);
[27] Cochran et al.\ (2004);
[28] Cochran et al.\ (2007);
[29] Collier-Cameron et al.\ (2007);
[30] Delfosse et al.\ (1998);
[31] Doellinger et al.\ (2007);
[32] Eggenberger et al.\ (2006);
[33] Endl et al.\ (2007);
[34] Fischer et al.\ (1999);
[35] Fischer et al.\ (2000);
[36] Fischer et al.\ (2002a);
[37] Fischer et al.\ (2002b);
[38] Fischer et al.\ (2003);
[39] Fischer et al.\ (2005);
[40] Fischer et al.\ (2006);
[41] Fischer et al.\ (2007);
[42] Frink et al.\ (2002);
[43] Galland et al.\ (2005);
[44] Gaudi et al.\ (2008);
[45] Ge et al.\ (2006);
[46] Griessmeier et al.\ (2007);
[47] Hatzes et al.\ (2005);
[48] Joergens \& Mueller (2007);
[49] Johns-Krull et al.\ (2007);
[50] Johnson et al.\ (2006a);
[51] Johnson et al.\ (2006b);
[52] Johnson et al.\ (2007a);
[53] Johnson et al.\ (2007b);
[54] Jones et al.\ (2001);
[55] Jones et al.\ (2002);
[56] Jones et al.\ (2006);
[57] Konacki et al.\ (2003);
[58] Konacki et al.\ (2004);
[59] Korzennik et al.\ (2000);
[60] Kovacs et al.\ (2007);
[61] Kurster et al.\ (2000);
[62] Latham et al.\ (1989);
[63] Lo Curto et al.\ (2006);
[64] Lovis \& Mayor (2007);
[65] Lovis et al.\ (2005);
[66] Lovis et al.\ (2006);
[67] Mandushev et al.\ (2007);
[68] Marchi (2007);
[69] Marcy \& Butler (1996);
[70] Marcy et al.\ (1999);
[71] Marcy et al.\ (2005);
[72] Marcy, Butler, \& Vogt (2000);
[73] Marcy, Cochran, \& Mayor (2000);
[74] Mayor \& Queloz (1995);
[75] Mayor et al.\ (2003a);
[76] Mayor et al.\ (2003b);
[77] McCarthy et al.\ (2004);
[78] McCullough et al.\ (2006);
[79] Melo et al.\ (2007);
[80] Mitchell et al.\ (2003);
[81] Moutou et al.\ (2005);
[82] Naef et al.\ (2001a);
[83] Naef et al.\ (2001b);
[84] Naef et al.\ (2003a);
[85] Naef et al.\ (2003b);
[86] Naef et al.\ (2007);
[87] Neuh\"{a}user et al.\ (2005);
[88] Niedzielski et al.\ (2008);
[89] Noyes et al.\ (2007);
[90] O'Donovan et al.\ (2006);
[91] O'Donovan et al.\ (2007);
[92] O'Toole et al.\ (2006);
[93] Pal et al.\ (2008);
[94] Pepe et al.\ (2002);
[95] Perrier et al.\ (2003);
[96] Pollaco et al.\ (2008);
[97] Pont et al.\ (2007);
[98] Queloz et al.\ (2000);
[99] Rasio (1994);
[100] Reffert et al.\ (2006);
[101] Robinson et al.\ (2007);
[102] Sahu et al.\ (2006);
[103] Santos et al.\ (2000);
[104] Santos et al.\ (2001);
[105] Santos et al.\ (2006);
[106] Santos et al.\ (2007);
[107] Sato et al.\ (2003);
[108] Sato et al.\ (2005);
[109] Sato et al.\ (2007);
[110] Sato et al.\ (2008);
[111] Setiawan (2003);
[112] Setiawan et al.\ (2003);
[113] Setiawan et al.\ (2005);
[114] Setiawan et al.\ (2007);
[115] Setiawan et al.\ (2008);
[116] Silvotti et al.\ (2007);
[117] Southworth (2008);
[118] Sozzetti et al.\ (2006);
[119] Tamuz et al.\ (2007);
[120] Tinney et al.\ (2000);
[121] Tinney et al.\ (2001);
[122] Tinney et al.\ (2002);
[123] Tinney et al.\ (2005);
[124] Torres et al.\ (2007);
[125] Udry et al.\ (2000);
[126] Udry et al.\ (2002a);
[127] Udry et al.\ (2002b);
[128] Udry et al.\ (2003a);
[129] Udry et al.\ (2003b);
[130] Udry et al.\ (2005);
[131] Vogt et al.\ (2000);
[132] Vogt et al.\ (2002);
[133] Vogt et al.\ (2005);
[134] Weldrake et al.\ (2008);
[135] Wilson et al.\ (2008);
[136] Wolszczan \& Frail (1992);
[137] Zucker et al.\ (2002);
[138] Zucker et al.\ (2004);
[139] da Silva et al.\ (2005).
}
\end{deluxetable}


}


We also carried out a targeted survey of some of the stars with
planets using \xmm\ (PI: V.Kashyap).  The stars were chosen from
the extreme ends of the distribution of $\scrA$, to provide a
contrast between stars with close-in planets and stars with
distant planets (see \S\ref{s:farout}).
These stars are listed in Table~\ref{t:xmmstars}.
We used the XMM Science Analysis System (SAS v7.0.0; 20060628\_1801)
to reduce the data and
obtained source counts within circular cells of radius $20''$
and a background estimated from nearby regions devoid of sources.
In the cases where no excess X-ray emission above the background
was detected at a significance of $\gtrsim0.997$ (corresponding
to a Gaussian-equivalent $3\sigma$ detection), count rate upper
limits were calculated as described by Pease et al.\ (2006).

\begin{deluxetable}{lcccccccc}
\tabletypesize{\scriptsize}
\tablewidth{0pt}
\tablecolumns{9}
\tablecaption{\xmm\ Data \label{t:xmmstars}}
\tablehead{
\colhead{Name} & \colhead{$\scrA$} & \colhead{ObsID/Revolution} & \multicolumn{3}{c}{{Exposure [s]}} & \multicolumn{3}{c}{{Count rate [ct ks$^{-1}$]}} \\
\colhead{} & \colhead{[AU]} & \colhead{} & \colhead{MOS-1} & \colhead{MOS-2} & \colhead{PN} & \colhead{MOS-1} & \colhead{MOS-2} & \colhead{PN}
}
\startdata
HD\,46375 & 0.041 & 0304202501/1071 & 8215.9 & 8464.2 & 6123.0 & $2.13 \pm 0.77$ & $<1.5$ & $11.2 \pm 1.9$ \\
HD\,187123 & 0.042 & 0304203301/1166 & 14586.7 & 15368.0 & 9382.7 & $<1.25$ & $<1.11$ & $<3.25$ \\
HD\,330075 & 0.043 & 0304200401/1037 & 15309.5 & 15772.6 & 13400.2 & $<0.8$ & $<0.8$ & $<1.7$ \\
HD\,217107 & 0.073 & 0304200801/0995 & 7536.3 & 7797.8 & 5497.1 & $<1.1$ & $<1.14$ & $4.1 \pm 1.5$ \\
HD\,130322 & 0.088 & 0304200901/1028 & 6295.6 & 6758.4 & 4208.7 & $4.4 \pm 1.1$ & $4.14 \pm 1.1$ & $19 \pm 2.7$ \\
HD\,190360 & 0.128 & 0304201101/0985 & 3078.0 & 3837.3 & 2573.9 & $<9.76$ & $<2.93$ & $<13.9$ \\
HD\,195019 & 0.138 & 0304201001/1167 & 10327.4 & 10259.4 & 8211.5 & $<1$ & $<1$ & $2.73 \pm 1.2$ \\
\multicolumn{9}{l}{\hfil} \\
47\,UMa & 2.11 & 0304203401/1191 & 8564.7 & 7735.4 & 6131.2 & $7.81 \pm 1.1$ & $8.1 \pm 1.2$ & $30.7 \pm 2.6$ \\
HD\,50554 & 2.38 & 0304203201/1163 & 1125.7 & 1117.0 & 1098.1 & $<6.3$ & $<4.4$ & $<7.9$ \\
HD\,216435 & 2.70 & 0304201501/1166 & 2045.0 & 2793.5 & 183.05 & $12.8 \pm 3.5$ & $8.21 \pm 2.4$ & $84.4 \pm 30.4$ \\
HD\,216437 & 2.70 & 0304201601/0979 & 4163.5 & 4396.5 & 2677.1 & $<2.3$ & $<2.15$ & $9.4 \pm 2.9$ \\
14\,Her & 2.77 & 0304202301/1059 & 4565.7 & 4570.9 & - & $<5.66$ & $<7.83$ & - \\
HD\,70642 & 3.30 & 0304201301/1159 & 13509.7 & 13489.1 & 10763.8 & $<0.9$ & $<0.9$ & $3.1 \pm 0.99$ \\
HD\,33636 & 3.56 & 0304202701/1054 & 10807.2 & 10847.9 & 6825.0 & $4.1 \pm 0.84$ & $4.2 \pm 0.85$ & $19.4 \pm 2.1$ \\
\enddata
\end{deluxetable}

Overall, we find that 70 of the stars in the sample
have been detected serendipitously or in
pointed observations (Table~\ref{t:egp_x}).
For those stars left undetected, we first
determine an upper limit from the \rass\ data, by 
estimating the number of counts required for a
$3\sigma$ detection (see Pease et al.\ 2006)
given the observed count rate
in the 0.1-2.4~keV band at the location of the source
and the accumulated exposure time 
(in the all-sky maps from the survey; Snowden et al.\ 1997).
If a pointed \xmm\ observation exists, we use an upper limit
derived from that observation.
We further impose a limit of
$\frac{L_X}{L_{bol}}<10^{-3}$ since stars do not exceed this limit on
average (though such limits may be exceeded on occasion when a large
flare occurs).
For the detected stars, we compute a nominal counts-to-energy conversion
factor, $cecf$ assuming a coronal source spectrum with similar
temperature components of 2 and 5 MK and an absorption column
of $10^{18}$ cm$^{-2}$.  For \xmm\ data, we computed the $cecf$
at a higher temperature (10 MK) to ensure that similar values of
fluxes are obtained for sources detected with both the MOS and
the pn detectors.
Adopting a single value of $cecf$ for all coronal sources observed with
a given detector introduces systematic errors of $\approx30$\%, comparable
to the statistical errors present in the measurements, and significantly
smaller than the intrinsic variability in the sources and other systematic
errors present in the the X-ray luminosity functions; the dominant
source of uncertainty in our results is sample variance, arising from
the intrinsic variations in the X-ray luminosity functions.
Using WebPIMMS~\footnote{
{\tt http://heasarc.gsfc.nasa.gov/Tools/w3pimms.html}}
v3.4 and XSPEC v10, we estimate that the $cecf$ for
\begin{eqnarray}
\nonumber ASCA/{\rm GIS} &=& 3~\times~10^{-10}  \,,\\
\nonumber Einstein/{\rm IPC} &=& 1.8~\times~10^{-11}  \,,\\
\nonumber EXOSAT/{\rm LE} &=& 6.5~\times~10^{-10}  \,,\\
\nonumber ROSAT/{\rm HRI} &=& 2.8~\times~10^{-11}  \,,~{\rm and}\\
\nonumber ROSAT/{\rm PSPC} &=& 6.5~\times~10^{-12} ~~{\rm ergs~cm^{-2}~ct^{-1}} \\
\nonumber XMM/{\rm EPIC-pn} &=& 1.9~\times~10^{-12} ~~{\rm ergs~cm^{-2}~ct^{-1}} \\
\nonumber XMM/{\rm EPIC-MOS} &=& 8.4~\times~10^{-12} ~~{\rm ergs~cm^{-2}~ct^{-1}} \\
\end{eqnarray}
in the 0.1-4.5 keV passband;\footnote{
There are slight differences in the conversion factors among different
catalogs from the same instrument, e.g., {\tt rassbsc} and {\tt wgacat}.
However, these discrepancies are much smaller than the uncertainty caused
by the source spectral distributions, and has no effect on our analyses.
}
in case a star has been detected multiple times with different instruments,
the flux measurement with the best S/N is reported in column 8 of
Table~\ref{t:egp_x}.  Similarly, if observed multiple times but never
detected, we report the lowest of the upper limits computed for that star.

\clearpage

\section{Analysis}
\label{s:analysis}

Here we carry out a series of statistical tests on the sample of
stars with close-in EGPs that were first detected, or at least
verified, using the radial-velocity method.
We first analyze the sample as a whole
(\S\ref{s:samplestat}), then search for correlations with
orbital radius $\scrA$ (\S\ref{s:adepend}), and finally search
for differences within extremal subsamples to increase the
contrast of the signal (\S\ref{s:farout}).

\subsection{Sample Statistics}
\label{s:samplestat}

Of the $>$230 stars in our sample thus far identified as possessing EGPs,
70 are found to be X-ray emitters.
If we exclude giants, we are left with a sample of
180 main sequence star systems with 46 X-ray detections.
This is a smaller fraction of X-ray
bright stars than is found for field stars in the solar neighborhood
(e.g., Schmitt 1997 finds 95\% of F stars and 83\% of G stars within
13 pc are detected with \rosat, while Maggio et al.\ 1997 find that
$\approx$50\% of G stars within 25 pc that were observed with \einstein\
are detected).

\begin{figure*}
\includegraphics[width=3in,angle=0]{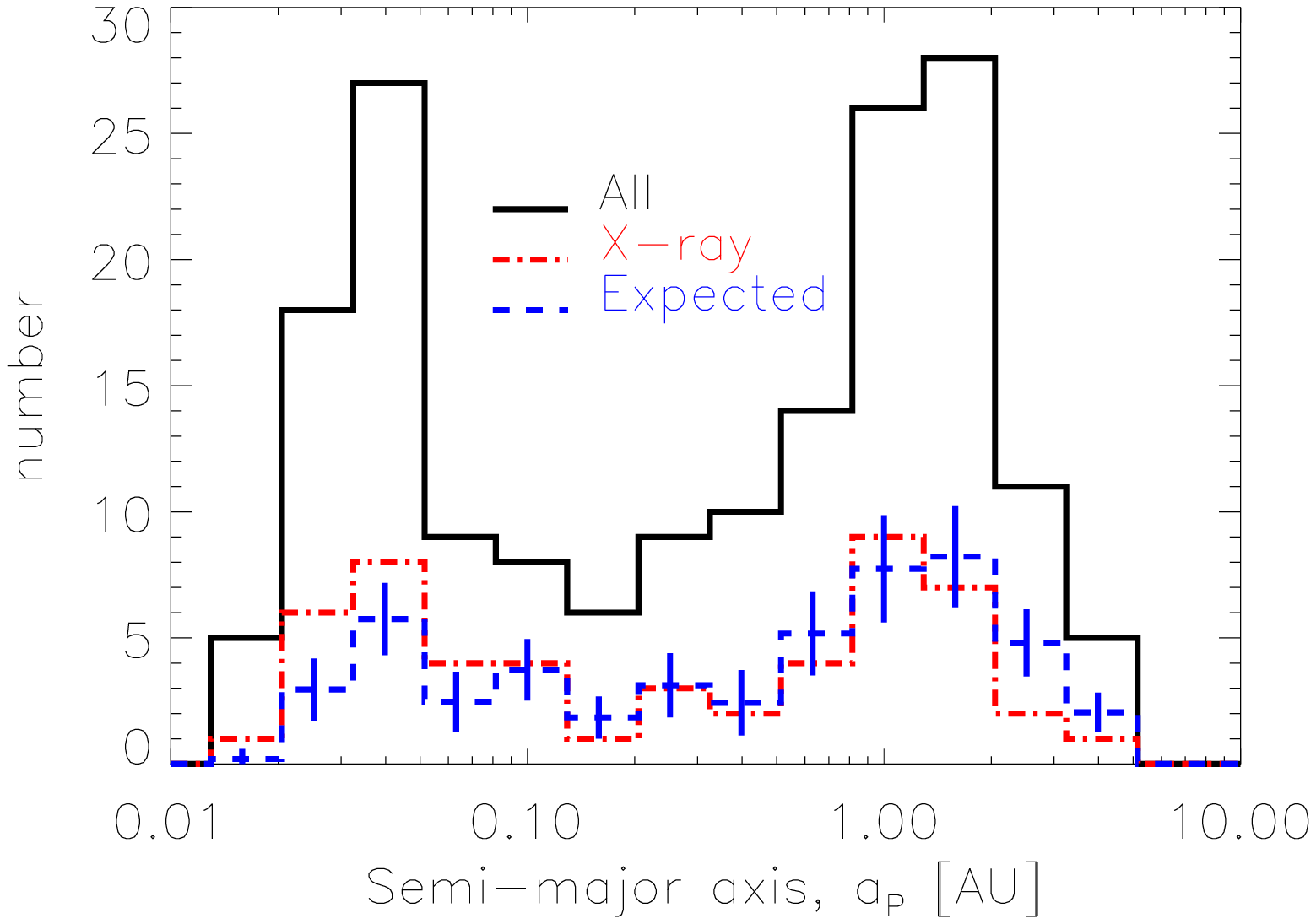}	
\includegraphics[width=3in,angle=0]{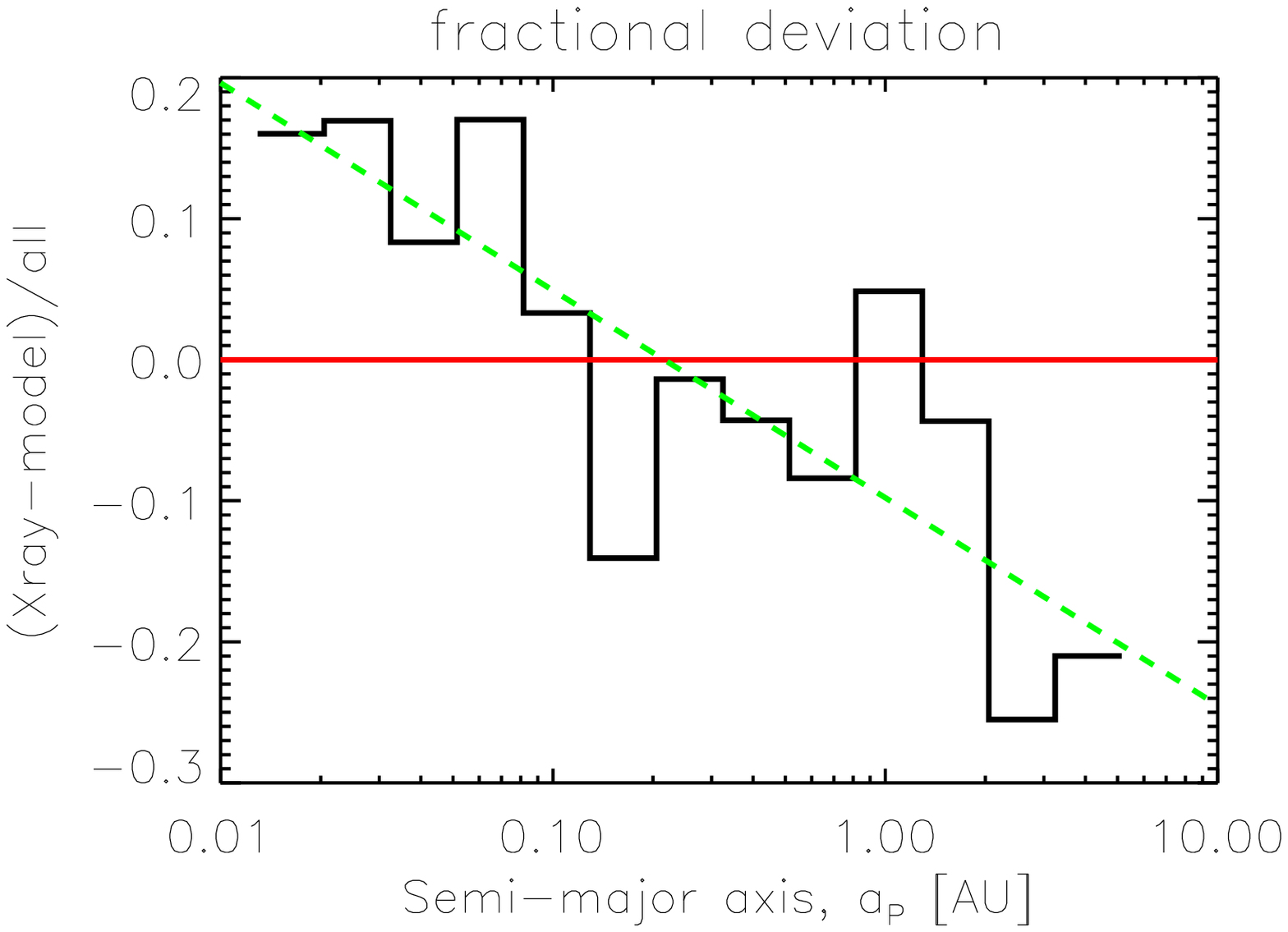}	
\caption{Distribution of observed and expected number of stars,
as a function of the semi-major axis, $\scrA$ of the closest planet.
{\sl Left:}
The solid histogram represents the sample of main sequence stars
among those listed in
Table~\ref{t:egp_x}, while the dot-dashed histogram
represents the subset of those stars which have been detected in X-rays.
The dashed histogram with error-bars denoted by vertical line segments
is the distribution generated by adopting known X-ray luminosity
functions and carrying out Monte Carlo simulations for stars at
the specific distances given in Table~\ref{t:egp_x}.
{\sl Right:}
The histogram shows the fractional difference between the predicted
and the actual number of X-ray detections in the sample.  The solid
horizontal line depicts the case of a perfect match, while the
dashed line shows the best-fit linear fit to the residuals.  This
shows that there may exist a weak trend in the data towards a higher
number of X-ray detections for stars that have close-in planets.
\label{f:numdist}}
\end{figure*}

A direct comparison of these surveys with our star sample is
however misleading; one must account for the intrinsic
variations in X-ray luminosities and the inhomogeneous scatter
in distances in our sample.  To investigate this,
we carry out a Monte Carlo simulation on our sample of stars,
fixing each star at its given distance, but allowing their X-ray
luminosities to vary.  These luminosities are obtained by adopting
the known X-ray luminosity functions (derived from statistically
complete samples of nearby field stars using \einstein\ data;
they thus include all the variations known to exist for
coronal sources; cf.\ Kashyap et al.\ 1992).
We generate 1000 realizations of
X-ray fluxes at Earth using these luminosity functions for each
star, and consider them to be detected in X-rays if the flux at
Earth is $>2 \times 10^{-13}$ ergs~cm$^{-2}$~s$^{-1}$ (corresponding
to the typical \rass\ sensitivity; Schmitt 1997).
The resulting frequency distributions of imputed detections can then
be compared to the observed frequency distribution of X-ray detections
in our sample.
This comparison is shown in Figure~\ref{f:numdist}, where the
frequency distributions of actual X-ray detections in our sample
(dash-dotted line) and of simulated X-ray detections for a
nominal set of field stars located at the same distances as
the stars in our sample (dashed line, with error bars derived
from averaging over the Monte Carlo realizations) are shown
as a function of the planetary orbit size.  Also shown are
the fractional residuals between the actual and predicted
number counts (solid histogram), which shows a weak trend
towards a higher efficiency of detection for stars with
close-in planets (dashed line; slope=$-0.1\pm0.03$).
A quantitative comparison between the two distributions yields
a reduced $\chi^2$ of $\approx1.8$, denoting that they are
marginally statistically
distinguishable.

\begin{figure}
\centerline{\includegraphics[width=4.5in,angle=0]{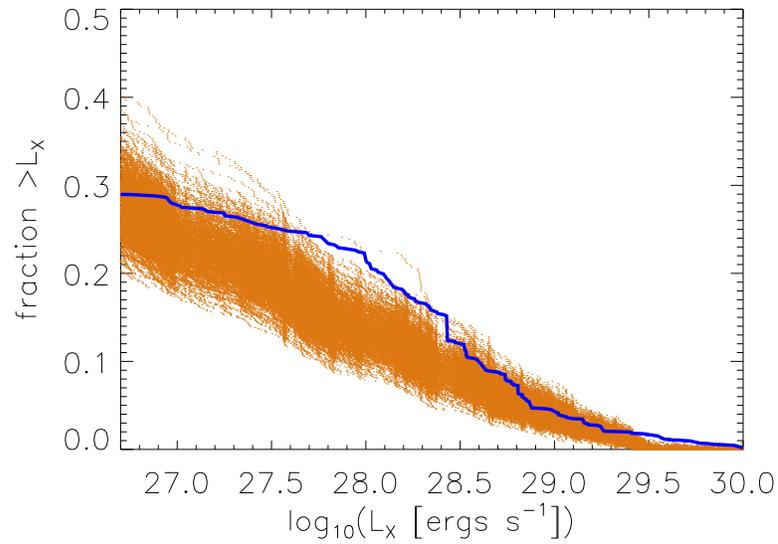}}	
\caption{X-ray luminosity distribution function.  The fraction
of stars with a given luminosity $L_X$ is shown for both the
main sequence sample from Table~\ref{t:egp_x} (solid histogram) as
well as for the expected distribution based on a model of the
Galaxy (shaded region).  The width of the shading represents
the statistical error present, both in terms of the numbers of
stars as well as the uncertainties inherent in the construction
of the X-ray luminosity functions.  The two cannot be statistically
distinguished under a Kolmogorov-Smirnoff test.
\label{f:xlf}}
\end{figure}

A comparison of the X-ray luminosity distribution of the sample
of stars with planets, compared to a nominally unbiased distribution
from field stars is shown in Figure~\ref{f:xlf}.
The expected X-ray distributions are shown as the shaded region,
and the actual luminosity distribution (derived as a Kaplan-Meier
estimator; see e.g., Schmitt 1985, Feigelson \& Nelson 1985)
as the solid histogram.
A formal Kolmogorov-Smirnoff test between the two shows that
the null hypothesis that the data are drawn from the full
sample cannot be ruled out.
Most of the differences can be attributed to detections over
a range of $L_X\sim3\times10^{27-28}$~ergs~s$^{-1}$.  While
not definitive, this again suggests the need for a more
sensitive analysis.

\subsection{Correlation with $\scrA$}
\label{s:adepend}

We show above (\S\ref{s:samplestat}) that, taken as a whole, the
set of main sequence stars with EGPs is similar in gross
characteristics to, though not identical to, the field star sample.
Given that, we next consider possible variations within our sample,
to test whether the EGPs have any measurable effect on levels of
X-ray emission.
Because frequency distributions are heavily model dependent and
are limited to integer values, they do not have sufficient
statistical power to detect weak variations in the properties
of the stars.
We expect that the effect of EGPs will increase as the orbital
distance decreases, and vice versa; we therefore check for
direct correlations of various sample parameters with the orbital
distance of the EGP.  

\begin{figure}
\centerline{\includegraphics[width=4.5in,angle=0]{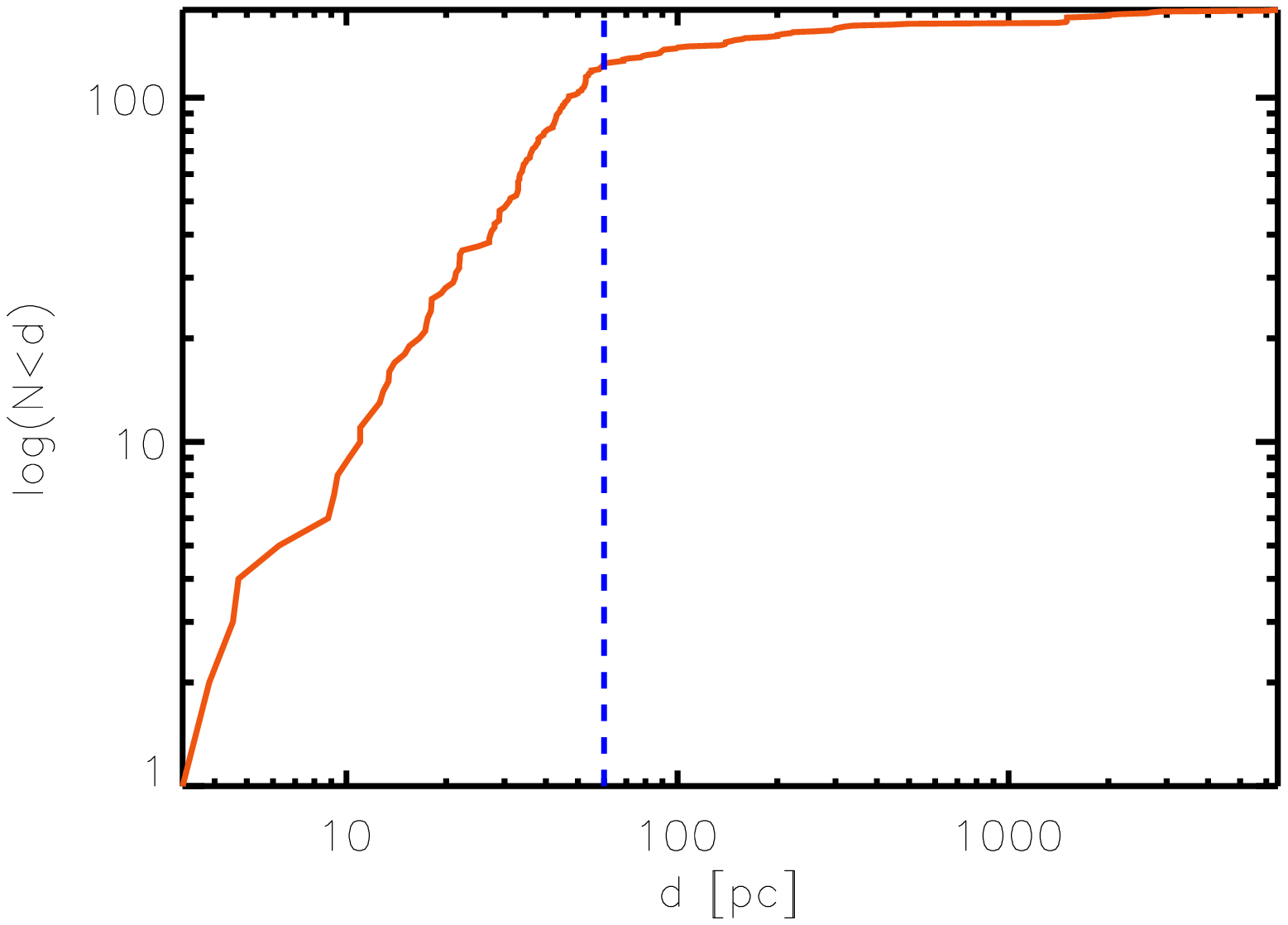}}	
\caption{The distribution of the number of main-sequence
star systems with distance.  The cumulative number within a
given distance is plotted as the solid curve.  The vertical
dashed line represents the distance limit of the statistically complete
subsample at 60~pc, within which the star systems are uniformly
distributed.
\label{f:logNlogd}}
\end{figure}

We also note that the sample is not volume limited, and that
this may introduce a number of complications into the analysis.
We therefore consider an additional filter to obtain a subset
of the main-sequence stars that are within 60~pc.  The number
distribution of these stars is
uniform within this distance (see Figure~\ref{f:logNlogd}), and this
subsample comes closest to a statistically complete volume
limited sample.

We expect {\sl a priori} that tidal and magnetic effects due
to close-in giant planets will manifest themselves as a trend
in stellar activity as a function of the planetary orbit $\scrA$.
We have thus carried out detailed statistical analyses to
measure correlations of the X-ray luminosity $L_X$ and
the activity indicator $\frac{L_X}{L_{Bol}}$ with $\scrA$.
We have carried out these tests for the full sample of main-sequence
star systems which have had planets detected via the Radial Velocity
method, as well as for the subsample of stars which are known
to be X-ray emitters, and for the subsample of stars which lie
within 60~pc.
If a given star has multiple planets detected, we choose the
planet with the smallest semi-major axis as defining $\scrA$.
We compute both Pearson's and Spearman's Rank Correlation
coefficients, and report the results in Table~\ref{t:acorrs}.
The errors on Pearson's coefficient are derived via bootstrapping
by sampling with replacement.\footnote{Where measurement errors
are available (e.g., for $L_X$), we include them in the simulations
by sampling with a Gaussian of standard deviation equal to the
measurement error and a mean equal to the maximum likelihood
value; upper limits are dealt with using a uniform distribution.}
The significance of Spearman's $\rho$ is denoted by the $p$-value,
which measures the probability that the observed value of $\rho$
can be obtained as a chance fluctuation.

\begin{figure}
\centerline{\includegraphics[width=4.5in,angle=0]{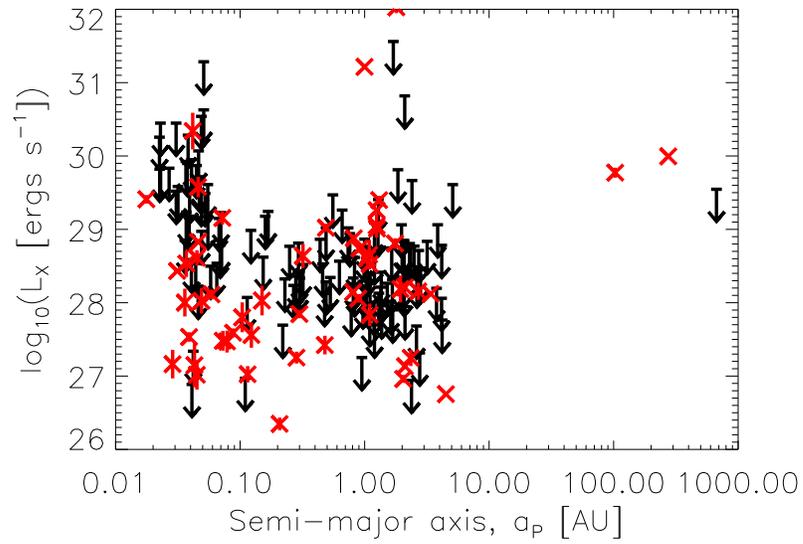}}	
\caption{X-ray luminosities of the sample stars as a function
of the orbital distance of the closest EGP.  The downward
arrows mark upper limits based on non-detections in \rass .
The X-ray detections are marked with an `x'; the errors on
the measured fluxes are denoted by vertical bars.
\label{f:alx}}
\end{figure}

\begin{deluxetable}{clll}
\tablewidth{0pt}
\tablecolumns{4}
\tablecaption{Correlation of orbital semi-major axis $\scrA$ with stellar parameters\tablenotemark{a} \label{t:acorrs}}
\tablehead{
\colhead{Parameter} & \colhead{Dataset} & \colhead{Pearson's Coefficient\tablenotemark{b}} & \colhead{Spearman's $\rho$\tablenotemark{c}}
}
\startdata
$L_X$ & Full sample & $-0.01 \pm 0.06$ & $-0.09 \pm 0.05$ \\
\hfil & X-ray detections & $-0.03 \pm 0.1$ & 0.01 ($p=0.96$) \\
\hfil & volume limited & $0.02 \pm 0.29$ & -0.08 ($p=0.38$) \\
\hfil & \hfil & \hfil & \hfil \\
$\log_{10}\left(\frac{L_X}{L_{Bol}}\right)$ & Full sample & $-0.04 \pm 0.1$ & $-0.09 \pm 0.05$ \\
\hfil & X-ray detections & $0.16 \pm 0.09$ & 0.006 ($p=0.96$) \\
\hfil & volume limited & $0.26 \pm 0.12$ & -0.08 ($p=0.38$) \\
\hfil & \hfil & \hfil & \hfil \\
$d_*$ & Full sample & $-0.19 \pm 0.03$ & -0.2 ($p=0.003$) \\
\hfil & X-ray detections & $-0.28 \pm 0.1$ & -0.07 ($p=0.6$) \\
\hfil & volume limited & $0.09 \pm 0.06$ & -0.07 ($p=0.5$) \\
\hfil & \hfil & \hfil & \hfil \\
$R_*$ & Full sample & $-0.023 \pm 0.045$ & -0.05 ($p=0.47$) \\
\hfil & X-ray detections & $-0.05 \pm 0.09$ & 0.10 ($p=0.48$) \\
\hfil & volume limited & $-0.05 \pm 0.1$ & 0.17 ($p=0.06$) \\
\hfil & \hfil & \hfil & \hfil \\
\enddata
\tablenotetext{a}{Stars from programs such as OGLE, SWEEPS, TrES,
WASP, etc., which initially detected planets via the photometric
transit method and not the radial velocity method, have been excluded
from these tests.}
\tablenotetext{b}{Pearson's linear correlation coefficient; errors
derived via bootstrapping by sampling with replacement.  Where
possible, measurement errors and upper limits are accounted for
via Monte Carlo simulations.}
\tablenotetext{c}{Rank correlation of two populations; the $p$ value
denotes the two-sided significance of its deviation from 0 by random
chance, i.e., small values indicate significant correlation.  Where
$p$-values are not quoted, the $1\sigma$ error on the correlation
coefficient, computed via Monte Carlo simulations, is shown.}
\end{deluxetable}

We show the distribution of X-ray detections and upper limits for
the full sample in Figure~\ref{f:alx} as a function of $\scrA$.
No trend is visually discernible here in $L_X(\scrA)$.  Detailed
correlation analysis (see Table~\ref{t:acorrs}) shows that there may
be a slight
negative correlation: for the full sample,
Spearman's Rank correlation coefficient is $\rho=-0.09\pm0.05$
(the $1\sigma$ error bar on $\rho$ is computed via 10000 Monte Carlo
simulations with the uncertainties in $L_X$ taken into account
at each iteration).
This is weakly significant, but the correlation becomes less
significant when smaller samples are considered (for the
X-ray detections, $\rho=0.01$, with $p=0.96$ even when
uncertainties in $L_X$ are ignored, and for the volume
limited sample, $\rho=-0.08$ with $p=0.38$).
We also compute
Pearson's correlation coefficient, which uses the information
of the actual values of the luminosities and not just their
rank order, and thus produces statistically more powerful
estimates.  As before, we obtain
error bars by carrying out Monte Carlo simulations
that take into account the uncertainties in $L_X$;
we adopt a Gaussian sampling function with $\sigma$ equal
to the measured error for detected sources and a flat
distribution in $\log{L_X}$ for undetected sources.
We find that Pearson's coefficient is $-0.01\pm0.06$ for the
full sample, $-0.03\pm0.1$ for the X-ray detected
sample, and $0.02\pm0.3$ for the volume limited sample.


A similar weak correlation is found when the
correlation of the activity indicator $\frac{L_X}{L_{Bol}}$ with
$\scrA$ is considered.  For the set of all stars, Pearson's coefficient
is $-0.04 \pm 0.11$ (Spearman's $\rho=-0.25\pm0.001$),
for only X-ray detected stars,
it is $0.16 \pm 0.1$ (Spearman's $\rho=0.01$; $p=0.96$),
and for the volume limited sample,
$0.26\pm0.12$ (Spearman's $\rho=-0.08$; $p=0.38$).
We consider the activity indicator in greater detail
below (see \S\ref{s:bias}) in establishing the magnitude of the
sample bias.

Since we do not expect that $L_X$ and $\scrA$ are linearly
related, and because the dynamic range is large, it is not
surprising that the correlation tests are contradictory and
inconclusive.  A more sensitive analysis is therefore
required, and we adopt a technique wherein the contrast in
the data is maximized (see \S\ref{s:farout}) by increasing
the lever arm between extremal subsamples.

In addition to the X-ray luminosity $L_X$, we also consider
the correlations of the stellar distance and stellar radius
with $\scrA$.  We expect a negative correlation to exist with
distance to the star, simply because it is easier to detect
planets and activity around nearby stars.  This is borne out
by the correlation analysis: we find that Pearson's coefficient
is $-0.19\pm0.03$ (Spearman's $\rho=-0.21$, with $p=0.003$)
for the entire sample, and for the X-ray detected sample
alone, Pearson's coefficient is $-0.28\pm0.1$ (Spearman's
$\rho=-0.07$, $p=0.6$).  For the volume limited sample, in
contrast, we obtain a Pearson's coefficient of $0.09\pm0.06$
(Spearman's $\rho=0.07$, $p=0.46$).
Note that here we determine the errors via bootstrapping by
sampling with replacement.
We expect to find no correlation with stellar radius, and
this expectation is borne out by the correlation analysis.
For the full sample, Pearson's coefficient is $-0.02\pm0.04$
(Spearman's $\rho=-0.05$, $p=0.47$); for the X-ray
detected sample, $-0.05\pm0.09$ ($\rho=0.1$, $p=0.48$);
and for the volume limited sample, $-0.05\pm0.1$ ($\rho=0.17$, $p=0.06$).

\subsection{Extremal Subsamples}
\label{s:farout}

\begin{figure}
\centerline{\includegraphics[width=4.5in,angle=0]{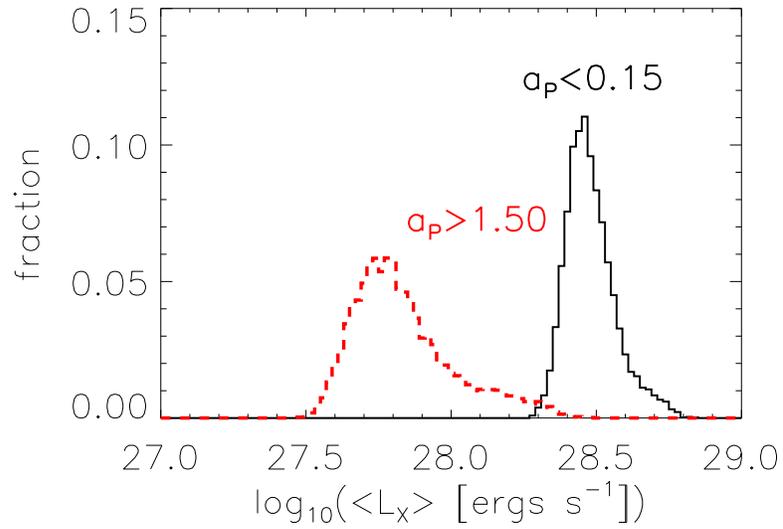}}	
\caption{Mean $L_X$ for the extreme samples.
The distribution of simulated means for the set of stars with
the closest giant planet having semi-major axis $\scrA<0.15$ AU is
shown as the solid histogram, and its counterpart for stars that
have giant planets with $\scrA>1.5$ AU is shown as the dashed histogram.
The overlap between the two distributions is $<1$\%.
\label{f:lohi}}
\end{figure}

\begin{figure}
\centerline{\includegraphics[width=4.5in,angle=0]{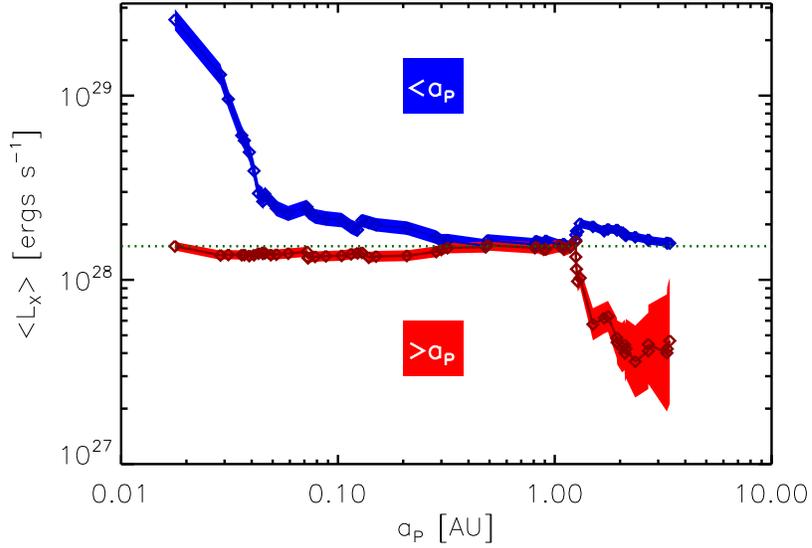}}	
\caption{Comparison of mean $L_X$ for extremal subsamples.  The
width of the shading represents the $1\sigma$ statistical error on the mean
$L_X$ derived via bootstrapping and Monte Carlo simulations (see
text).  The upper shaded region represents subsamples that include
close-in planets, $<L_X(\scrA<\scrA_{close-in})>$ and the lower
shaded region, the subsamples that include distant planets
$<L_X(\scrA>\scrA_{distant})>$.  The mean values in both cases are
shown as thick dark lines marked with diamonds within the shaded regions.
The distributions in Figure~\ref{f:lohi} correspond to vertical cuts
across the shaded regions here, at $\scrA_{close-in}=0.15$ (upper
region, for $<L_X(\scrA<0.15)>$) and $\scrA_{distant}=1.5$ (lower
region, for $<L_X(\scrA>1.5)>$).
\label{f:lohirun}}
\end{figure}

In order to clarify the variations in the X-ray activity statistically,
we consider the extreme cases of stars with planets that are close
to the primary (the ``close-in'' sample) and compare their luminosities
with those of stars with planets at large distances (the ``distant''
sample).\footnote{We are precluded from directly comparing our sample with
that of an independent sample of field stars because at this stage
it is not possible to be certain that the putative field star sample
is devoid of close-in EGPs.}
We choose these subsamples from the volume limited sample, i.e.,
main sequence star systems which are within 60~pc.
This signal is obviously dependent on the contrast between the
close-in and the distant samples, and in order to
obtain the best contrast it is necessary to choose samples that
are as far apart in their range of $\scrA$ as possible and yet
contain as large a sample of stars as possible.  A comprehensive
investigation of the best such separation is not feasible for
this sample because of the large number of X-ray non-detections
and the variety of tests that we carry out on the data.  However,
for our purposes it is sufficient to determine whether there
exists subsamples which show the requisite contrast, at some
separations, even if it is not necessarily the optimal one.
(Note that we do consider below the effect of varying the sizes
of the samples.)
We therefore adopt an {\sl ad hoc} separation based on the
population distribution (Figure~\ref{f:numdist}): we choose
$\scrA<0.15$~AU (corresponding to a dip in the frequency distribution
of stars as a function of orbital distance $\scrA$)
for the close-in sample, which results in a sample size of 40 stars,
20 of which are detected in X-rays.  By limiting the lower
bound of the distant sample such that there is separation of
an order of magnitude difference in $\scrA$ between the limits
of the two samples, we set for the distant sample, $\scrA>1.5$~AU;
8 of 38 stars in this subsample are detected in X-rays.
This range $\scrA$ has the additional advantage that similar
numbers of stars are found in each set.
The mean X-ray luminosities for these two subsamples are
found to be well separated, with the close-in sample being
significantly X-ray brighter (see Figure~\ref{f:lohi}).
We comment on alternative choices further below (see
\S\ref{s:activity}) and explicitly show (Figure~\ref{f:lohirun})
that this specific choice does not affect our results.

Because of the large number of upper limits in the dataset, we
cannot carry out simple hypothesis tests to verify whether these
two samples are derived from the same parent distribution.  Instead,
we carry out Monte Carlo realizations of the sample
as before (\S\ref{s:adepend}), using a Gaussian
error distribution with measured errors for the detected stars
and a flat uniform distribution in the log-scale
for the undetected stars.  We compute sample means for each set
of realizations, and the distribution of these means allows us
to determine whether the two samples are similar or different.\footnote{
Note that because fewer than half the stars in the sample
are detected in X-rays, we are precluded from using sample
medians as a summary estimator of the subsamples; the median
is not a robust estimator in this case.}

We find that the mean X-ray luminosity\footnote{Here
and henceforth the enclosing brackets ``$<>$'' denote the
mean value of a quantity; note that within these brackets we
often have numerical conditional expressions such as ``$<$''
(less than) and ``$>$'' (greater than), enclosed within
parentheses.}
of the close-in sample is
$<L_X(\scrA<0.15)>=10^{28.49\pm0.09}$~ergs~s$^{-1}$, and that of
the distant sample is $<L_X(\scrA>1.5)>=10^{27.85\pm0.18}$~ergs~s$^{-1}$.
The mean level of emission for the close-in sample is significantly
greater than for the distant sample, and as shown in Figure~\ref{f:lohi},
the two samples are found to be different at $>99$\% confidence
level.\footnote{
Note that the nearby strong X-ray source $\epsilon$\,Eri is in the
distant subsample.  A planet with a period $\sim10$ yr was detected
around it by Campbell, Walker, \& Yang (1998), but this detection
remains controversial due to the large intrinsic scatter of
$\sim 20$ m s$^{-1}$ in the radial velocity curves
(see Marcy et al.\ 2002;
{\tt http://exoplanets.org/esp/epseri/epseri.shtml}).
Excluding $\epsilon$\,Eri would decrease the mean $L_X$
of the distant sample even further, increasing the separation
between the two distributions.
}

As discussed above, our choice of the limiting $\scrA$ for the
two subsamples is {\sl ad-hoc}.  We have therefore considered
the effect on our result of of varying the subsample bounds of $\scrA$.
The mean $L_X$ for various subsamples is obtained using Monte Carlo
simulations as above, for various ranges of $\scrA$, and are shown
in Figure~\ref{f:lohirun}.  Two types of subsamples are considered:
one that includes all stars with planetary orbital radii $<\scrA_{close-in}$,
from which the mean X-ray luminosity $<L_X(\scrA<\scrA_{close-in})>$ is obtained
(upper shaded curve),
and another that includes all stars with orbital radii $>\scrA_{distant}$,
which results in $<L_X(\scrA>\scrA_{distant})>$ (lower shaded curve).  As can be
seen, the sample that includes close-in planets is invariably
more X-ray intense than the sample that includes distant planets
for almost all possible choices of $<\scrA_{close-in}$ and $>\scrA_{distant}$.
Note that these results do not change in any qualitative way if the
dM stars (which contribute significantly to the high mean $L_X$
point at small $\scrA$) are excluded from the sample.

We caution that this {\sl observed} difference in the mean $L_X$
between stars with close-in and distant planets cannot be
entirely attributed to the effect of close-in EGPs; there
are selection biases inherent in the sample that must be
accounted for.  By considering indirect
sample ensemble properties, we argue below (see \S\ref{s:bias})
that the bias inherent in our selected sample has a small
effect.
Based on physical grounds (see \S\S\ref{s:intro},\ref{s:binary}),
we expect that close-in giant planets could have a significant
effect on the X-ray activity level of the primary,
and the trend seen in the activity level as a function
of the size of the planetary orbit (semi-major axis $\scrA$) must
in large part be due to the effect of the close-in giant planet.

%
%

\section{Discussion}
\label{s:discuss}

\subsection{Sample Bias \label{s:bias}}

The sample of candidate stars for which planet searches
are conducted is subject to some subtle biases.  Some of
these biases have the effect of masking the signature of
planet induced activity and are difficult to quantify since
the planet detection processes are numerous, the programs are
still incomplete, and the rates of false positives and
false negatives are unknown.  We may however determine the
approximate extent of these biases by studying the ensemble
properties of the sample of stars with detected planets.
Here we describe these biases, and consider their effect on
our ability to detect intrinsic trends in X-ray activity.
Furthermore, because the set of EGPs is dominated by
those identified with the radial velocity method, we
shall limit our sample to those stars which have been
verified to have planetary systems by this method, and
will thus consider only those biases introduced by that
method.

\begin{enumerate}

\item {\bf Spectral Homogeneity}
Most of the stars for which planets are searched
for are solar like.
This is advantageous to our study since the stars considered
here are relatively homogeneous; therefore effects that
arise due to changes in the convective turnover timescales
leading to changes in the nature of the magnetic dynamo
at the high- and low-mass ends of the coronal sequence,
or due to the changing evolutionary states of the systems
may generally be ignored.
We further homogenize the sample by limiting it
to main sequence stars within 60~pc
(additional types of filtering to further homogenize
the samples has no effect on our results; see \S\ref{s:activity}).

Because the sample of stars is homogeneous, we do not expect
any correlations between the stellar radii $R_*$ and $\scrA$
to be present, and indeed we find that the data are consistent
with these expectations (Table~\ref{t:acorrs}).  We have also
carried out a full analysis of the extremal subsamples for
various subsets of the full dataset (see Figure~\ref{f:lohiratmany})
and show that our results are robust to such selections.
We thus conclude that in our sample there is no bias present
due to stellar type or size effects.

\item {\bf Stellar Distance}
For a given planetary mass, the detectability of
planets decreases as the distance of the planet from
the star ($\scrA$) increases, since the radial velocity
amplitude decreases.  Since detectability in general
decreases with distance to the star ($d_*$), we expect
to find more stars with small $\scrA$ at larger $d_*$.  In the
sample of stars with detected planets, we therefore expect
that $\scrA$ and $d_*$ are anti-correlated.  Indeed, we find
a slight negative correlation between $\scrA$ and $d_*$,
though it is weak (see Table~\ref{t:acorrs}).
This bias exists simply as a result of the
limitations of measurement statistics and is independent
of any stellar activity effects.  Thus, if we assume
that X-ray properties of stars are independent of the
distance limit, the only effect on the X-ray data will
be to have a larger fraction of higher upper limits
at smaller $\scrA$ (due to the larger numbers of more
distant stars).\footnote{
In our correlation analyses we have tested for the effect
that these higher upper limits may have by using uniform
distributions (bounded above by the value of the upper
limit) in both log space and normal space to describe
the censored data and find no qualitative difference
between the two cases.
}
Thus, stellar distance should have no effect on the
ensemble properties of the {\sl X-ray} luminosities of
stars from such a sample.  In any case, we carry out
most of our analysis with the volume limited sample,
which is explicitly designed to remove any such effect.

\item {\bf Selecting for reduced activity}
The process of planet detection via the radial
velocity method (RV; e.g., Butler, et al.\ 1996) is limited by the
amount of intrinsic RV jitter that may be caused by
stellar activity (Saar \& Donahue 1997, Saar et al.\ 1998;
see also Baluev 2008),
and hence planets are preferentially detected when they
are close-in, massive, and the primary star is not active.

Because radial velocity signatures are hard to
detect around active stars, there is a tendency to select
candidate stars for planet search {\sl against} activity.
Thus, {\sl a priori} there is an expectation that the
sample of stars with detected planets would have lower
activity levels than field stars of the same type.  But
as shown in \S\ref{s:samplestat}, the X-ray emission
from these stars is by and large consistent with field star emission
levels.  In any case, in order to avoid introducing
accidental biases by comparing the sample of stars with
EGPs to those without detected EGPs (note that a lack of
detection does not preclude the existence of planets
around the star -- even if the star is part of the
planet search program, there may be a planet that is
less massive or more distant than the sensitivity that
has been reached), we have chosen to compare the extrema
of a single distribution, viz., the sample of stars known to
possess giant planets.  Thus any global selection effect
that may exist in the sample towards lesser activity will
apply equally to both subsamples and its effects are
irrelevant for this study.

Also note that in some types of planet detection methods,
persistent stellar activity fixed to an active longitude
can mimic the signature of a planet (Lanza et al.\ 2008).
This results in false detections of planets at the stellar
orbital period.  However, this has an effect on photocentric
methods and have no effect on radial velocity or transit
methods currently in use.

\item {\bf Intra-sample trend in inhibition of activity}
If stellar activity inhibits planet detection, then it
will inhibit it selectively.  Planets at large $\scrA$ are
preferentially detected around stars with weak activity,
since high activity may mask the RV signal of the distant
planet.  Thus, in a sample of stars selected based
on the existence of planets, those stars with distant
planets are presumptively less active.  This bias thus
produces the same signature in activity trends that we
search for (see \S\ref{s:farout}), and thus interferes
with our study.

How can this bias be quantified?  Consider that the cause
of the bias is the excess jitter in velocity induced by
magnetic activity, which is related to the energy deposited
in the corona, which in turn is tracked by the ratio\footnote{
$\frac{L_X}{L_{bol}}$ is preferred for this bias measurement
because it is surface area independent.
}
of X-ray to bolometric luminosity $\frac{L_X}{L_{Bol}}$.  This
jitter serves to mask the RV signal, which is due to wobble
from gravitational effects and is therefore inversely
proportional to the orbital semi-major axis of the planet.
Thus, the bias will manifest itself as a positive correlation
between the formally independent parameters $\frac{L_X}{L_{Bol}}$
and $\frac{1}{\scrA}$.  Note that this correlation
is not necessarily linear because it is dependent on the observation
process, the cumulative observation time, etc.  Any existing
correlation also does not
imply causality, and indeed the velocities that are generated
by these two processes are quite different.  Nevertheless,
we expect that stars with large values of $\frac{L_X}{L_{Bol}}$
will primarily be present in our sample provided they are also
accompanied by small values of $\scrA$.  Thus, a measurement
of the correlation between these two variables serves as a
measurement of the sample bias.  For the sample of main sequence
stars within 60~pc that have EGPs detected via the RV method,
we find that
\begin{equation}
\label{e:vel2bias}
\log_{10}\frac{L_X}{L_{Bol}} = (-5.65 \pm 0.04) + (0.22 \pm 0.05) \cdot \log_{10}\frac{1}{\scrA} \,,
\end{equation}
where $\scrA$ is in units of [AU].
Here the errors are derived via Monte Carlo
bootstrapping on the X-ray flux measurement errors.  Thus,
we find that there does exist a weak, but statistically significant,
correlation between $\frac{L_X}{L_{Bol}}$ and $\frac{1}{\scrA}$.
This trend is shown in Figure~\ref{f:vel2}.  {\sl Conservatively
assuming that all of this correlation is attributable
to the inherent bias}, we can then estimate its effect
on the observed activity trend.  That is, for each
of the stars in the close-in and distant samples
(in \S\ref{s:farout}), given $\scrA$ and $L_{Bol}$ we
can compute a predicted $L_X(\scrA,L_{Bol})$ that can
then be used in the place of the measured $L_X$ to
carry out the same analysis.  The result of this
calculation is shown in Figure~\ref{f:velohi}, which
shows that the close-in and distant samples differ
by $\approx{1}\sigma$, and account for only a factor of
$\sim 2$ of the {\sl observed} difference between
the two samples.
It is important to note that the magnitude of this
bias will vary for every subsample, and that the
magnitude and direction of the bias will be case
specific.

\begin{figure}
\centerline{\includegraphics[width=4.5in,angle=0]{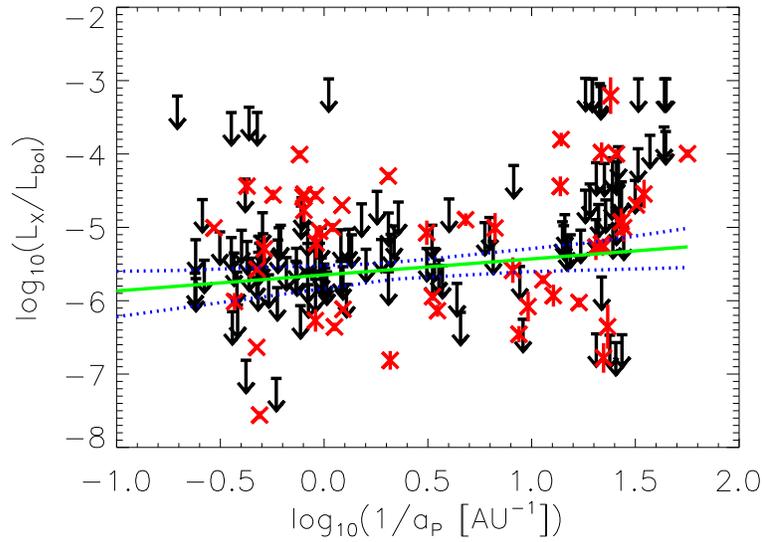}}	
\caption{The distribution of $\log\frac{L_X}{L_{Bol}}$ with
$\log\frac{1}{\scrA}$.
The X-ray detections are marked with `x's, with error bars denoted
with vertical lines; the X-ray non-detections are shown with downward
arrows.  Also shown are the best-fit regression line (solid curve)
and the envelope bound of the bootstrapped regression curves (dotted
lines).  Since more active stars have higher radial velocity jitter,
we expect fewer planets to be detected around stars at high
$\frac{L_X}{L_{Bol}}$ unless they also have high $\frac{1}{\scrA}$;
this correlation tracks an important sample bias (see text).
This trend is removed from the rest of the analysis.
\label{f:vel2}}
\end{figure}

\begin{figure}
\centerline{\includegraphics[width=4.5in,angle=0]{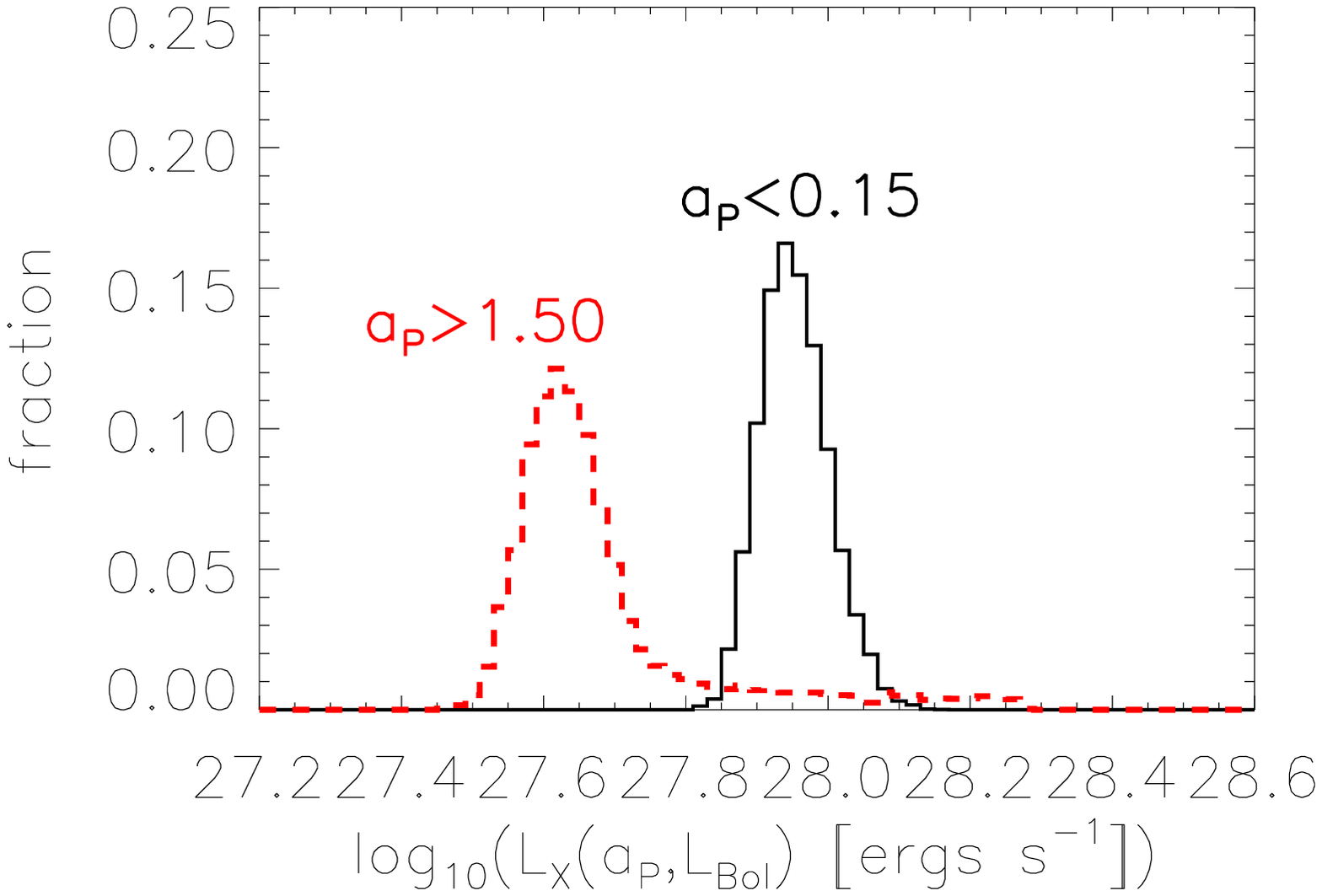}}	
\caption{Same as Figure~\ref{f:lohi}, but using $L_X(\scrA,L_{Bol})$
estimated from the regression analysis on $\frac{L_X}{L_{Bol}}$
with $\frac{1}{\scrA}$ (Equation~\ref{e:vel2bias}).
The two distributions differ by $\approx{2}\sigma$.
\label{f:velohi}}
\end{figure}

Note that $\frac{L_X}{L_{Bol}}$ is the preferred proxy for activity
because it is surface area independent.
However, because there are no $R_*$ biases in the sample, $L_X$ and
$\frac{L_X}{L_{Bol}}$ are similarly correlated with $\scrA$.  Thus,
a priori, we expect that accounting for the selective inhibition bias
of Equation~\ref{e:vel2bias} in the $\scrA$ vs $L_X$ dataset should
result in a complete removal of any luminosity difference between
the extremal samples, i.e., that
$\frac{<L_X({\rm close-in}>}{<L_X{\rm distant}>}\approx1$.
That this is manifestly not so (Figure~\ref{f:lohirat}) is an
indication that the differences in the distribution of $L_X$
for the extremal samples is systematic, and an inherent property
of the observed sample.

Furthermore, the bias measurement is conservative, i.e.,
we include some of the desired signal within the estimate of
the bias, and thus we {\sl overestimate} the magnitude of the
bias.  Because this bias is removed from the observed ratios
of the mean luminosities, our results provide a conservative
estimate of the true magnitude of the effect of planet-induced
activity.

\item {\bf Intrinsic sample variance}
Other parameters known to affect the level of stellar
activity are age, rotation, the Rossby number, metallicity,
temporal variability, etc.

Stellar ages are not
known with sufficient accuracy for us to consider them here,
and it is possible that there exists a trend where $\scrA$
is correlated with stellar age.  But note that such hidden
variables do not affect the results derived here, and simply
point to a more complex explanation of the connection between
close-in planets and stellar activity.

It is also as yet unclear what
type of effect a massive Jupiter-type planet will have
on these parameters.  For instance,
Rotational synchronization
by tidal interactions may be thought to increase $v\sin{i}$
and thereby coronal activity for close-in EGPs compared to
distant EGPs.  However, not only is this as likely to spin-down
a fast rotating star as to spin-up a slow-rotator, but also
the timescales for the synchronization of the stellar rotation
period with the planetary orbital period are very large to
begin with, and increase as $\scrA^6$ for distant planets
(see e.g., Drake et al.\ 1998 and references therein),
implying that most stars in the sample are {\sl not} synchronized
to the orbits of their close-in giant planets.
We therefore conclude that both theoretical
and observational prejudice points to evenly scattered values
for these stellar parameters in our sample.

\end{enumerate}

\subsection{Activity Enhancement}
\label{s:activity}

\begin{figure}
\centerline{\includegraphics[width=4.5in,angle=0]{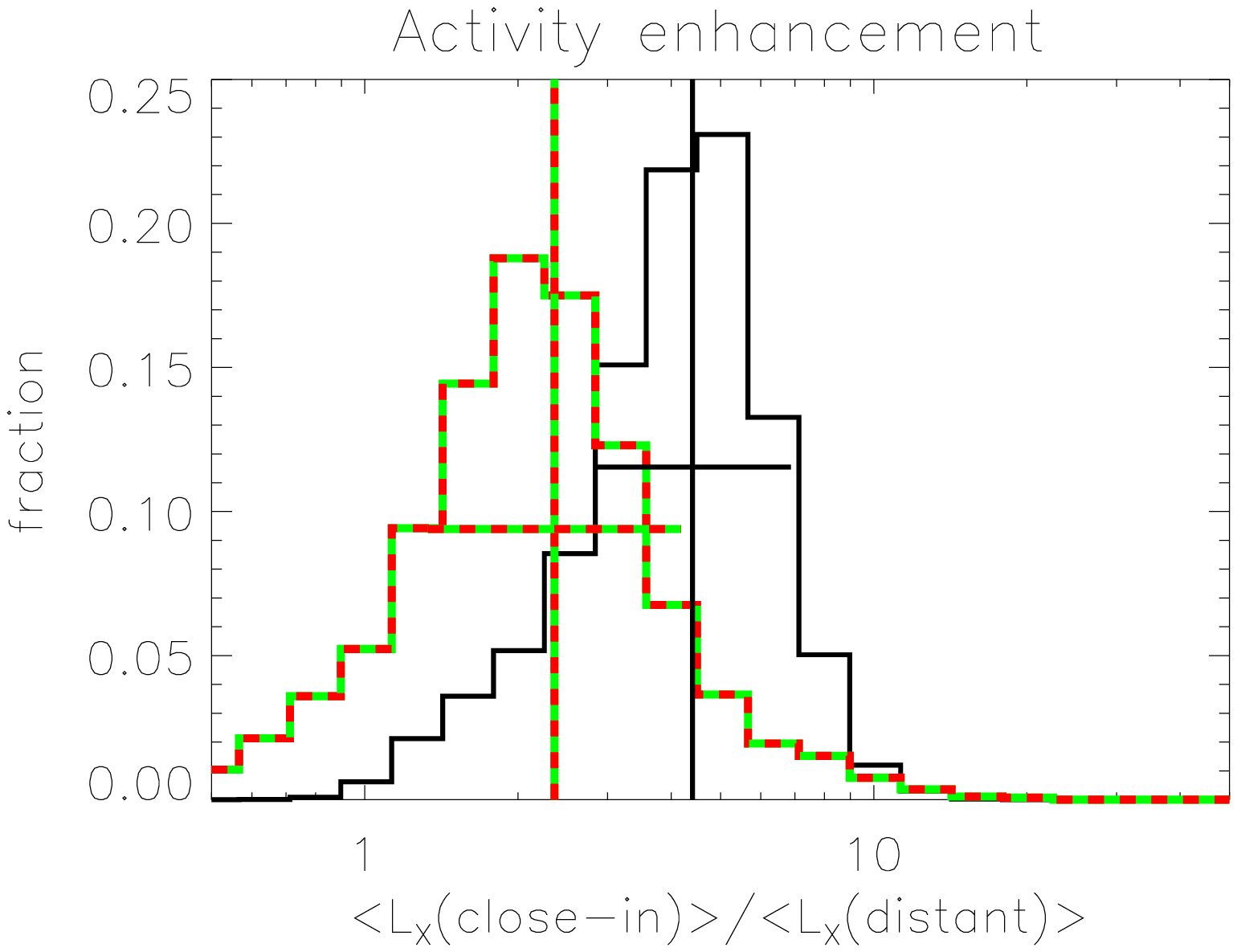}}	
\caption{Distributions of $<L_X{\rm (close-in)}>/<L_X{\rm (distant)}>$.
The distribution of the ratios of simulated $L_X$ for the close-in
($\scrA<0.15$ AU) and distant ($\scrA>1.5$ AU) samples is shown as the
solid histogram.  The same ratios, modified to account for the
sample bias (see \S\ref{s:bias}) are shown with dashed lines.
The means and standard deviations of the two distributions are
denoted with vertical and horizontal lines respectively (solid
and dashed, for the two distributions).  The data show that the
two extremal samples differ by a factor $\approx{4}$ (between
2.8 to 7 times), and after sample bias is accounted for, by
a factor $\approx{2}$ (between 1.3 and 4 times).
\label{f:lohirat}}
\end{figure}

We use the parameterization of the bias described above (see
\S\ref{s:bias}) to determine the residual signal, i.e., the
magnitude of the excess X-ray activity in the close-in subsample
compared to the distant subsample that can be attributed to the
effect of EGPs.  In order to simplify the calculation, we
consider the ratio of the average luminosities of the close-in
and distant subsamples calculated during the Monte Carlo simulation
described above (see analysis in \S\ref{s:farout}).  The
data from Figure~\ref{f:lohi} are shown as the solid histogram
in Figure~\ref{f:lohirat}, and confirm that the close-in
subsample is on average more X-ray luminous by a factor
$4^{<7}_{>2.8}$, where the bounds on the number indicate
the extent of the asymmetrical $1\sigma$-equivalent errors.
After correcting for the sample bias (by
dividing the original ratio by the ratio of the luminosities
as predicted by Equation~\ref{e:vel2bias}), we find that the
bias-corrected luminosity ratio decreases, but is $>1$ at
a significance of $0.95$.  This constitutes a definite detection
of activity enhancement, by a factor $2.3^{<4}_{>1.3}$
(dashed histogram in Figure~\ref{f:lohirat}).
This remaining enhancement can be
attributed to the effect of close-in giant planets on their
parent stars.
We therefore conclude that the close-in and
distant samples indeed show differing X-ray activity levels
and that a factor $\sim 2$ can be attributed to the effect
of close-in EGPs.
Note that the existence of this residual
enhancement does not in itself allow us to conclude that
there is a causal connection between the closeness of giant
planets to their primary stars and the activity levels on
those stars.  However, in conjunction with the Ca\,II~HK
enhancements observed by Shkolnik et al.\ 2003), our results
suggest that such a connection may be present.

\begin{figure}
\centerline{\includegraphics[width=4.5in,angle=0]{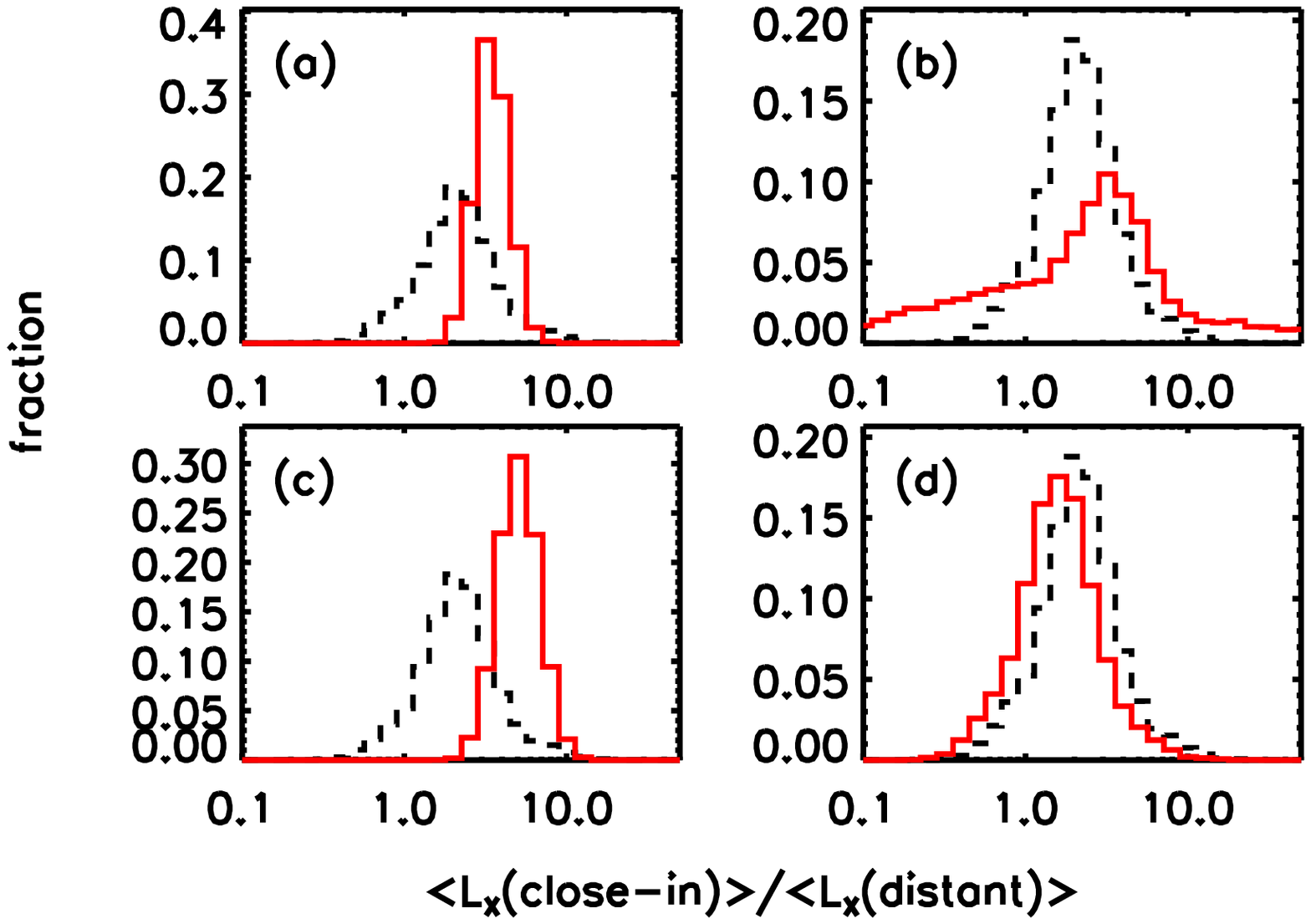}}	
\caption{Effect of variations of sample composition on the distribution
of $<L_X{\rm (close-in)}>/<L_X{\rm (distant)}>$.  The ratios of
$<L_X(\scrA<0.15)>/<L_X(\scrA>1.5)>$ are compared for different
representative subsets of the full sample of stars with detected planets.
In all plots, the baseline comparison from Figure~\ref{f:lohirat}, for
the main-sequence sample within 60~pc, is shown as the dashed
histogram.  The solid histograms are:
{\sl (a)} all X-ray detections,
{\sl (b)} the full sample of all stars with EGPs detected via the RV method,
{\sl (c)} the nearby main-sequence sample excluding dM stars, and
{\sl (d)} the full sample as in (b), excluding giants.
In all cases, the bias correction (see \S\ref{s:bias}) is done
separately for each sample.  In all cases, the evidence is
strong that stars with close-in giant planets are more active
than those with distant planets.
\label{f:lohiratmany}}
\end{figure}

As a further check on the sensitivity of our analysis to the
samples used, we compute the same ratio as in Figure~\ref{f:lohirat}
for a number of different subsamples.
The resulting distributions of the ratios
of mean luminosity for close-in versus distant subsamples are shown
in Figure~\ref{f:lohiratmany}; in all cases the close-in subsamples
have larger $<L_X>$ than the distant subsamples. 


Note that the above result is dependent on the bounds chosen
for $\scrA$ for both subsamples.
Clearly, the contrast between the two extremal sets would be heightened
if the range of $\scrA$ defining the two subsamples is shrunk further,
and vice versa.  This must also be balanced by the decreased robustness
of the result due to the consequent dearth of X-ray detections in the
subsamples.
A rigorous estimate of the ``best'' split between the close-in and
the distant samples is not feasible because of the large number of
stars that remain undetected in X-rays and the need to compute the
bias factor (Equation~\ref{e:vel2bias}) separately for each case.
Nevertheless, as we have shown above (Figure~\ref{f:lohirun}) there
exists a persistent difference in the mean luminosity between the
close-in and distant subsamples, and it favors the suggestion that
stars with close-in planets tend to be more active.
The best way to improve this result is to obtain
X-ray detections of more stars at the extreme ranges of $\scrA$.


\subsection{Dependence of the X-ray excess on $\scrA$}
\label{s:binary}

For tidally interacting close binaries in a sample
that included RS\,CVn stars, Schrijver \& Zwaan (1991)
found that the X-ray surface flux $F_X \propto \sqrt{a_*}$,
where $a_*$ is the distance to the cooler component.
Comparing the average X-ray luminosities
for close-in and distant subsamples that we derive above
(see \S\S\ref{s:farout},\ref{s:activity}), after sample
biases have been corrected for, we find that
\begin{equation}
<F_{X_{\rm corr}}> ~~ \propto ~~  <\scrA>^{-0.35 \pm 0.17} \,,
\end{equation}
which is a smaller dependence than that found by Schrijver \& Zwaan,
but the large error bar prevents a definitive conclusion.

It is well known that binary stars are generally
more X-ray active than single stars of the same type
and rotation rate
(see, e.g., Zaqarashvili, Javakishvili, \& Belvedere
2002 and references therein).  Numerous mechanisms have
been proposed to account for this so-called ``overactivity,''
generally based on tidal and magnetic interactions.  These
studies were extended to the case of stars with EGPs
and brown dwarfs by Cuntz et al.\ (2000).
They suggested that the energy generation due to tidal
interactions is proportional to the gravitational
perturbation
\begin{mathletters}
\begin{equation}
\frac{\Delta g_*}{g_*} = \frac{M_P}{M_*} \frac{2 R_*^3}{(\scrA-R_*)^3}
\end{equation}
where $M_*$ is the mass of the star and $M_P$ that of the EGP, and
the height of the tidal bulge,
\begin{eqnarray}
\nonumber h_{tide} &\propto& \frac{\Delta g_*}{g_*} R_* \\
\nonumber &\propto& \frac{M_P}{M_*} R_*^4 \scrA^{-3} \\
&\propto& \scrA^{-3} \,.
\end{eqnarray}
\end{mathletters}
Saar et al.\ (2004) estimated the energy released via reconnection
during an interaction of the planetary magnetosphere with
the stellar magnetic field,
\begin{eqnarray}
\nonumber F_{int} &\propto& B_* B_P v_{rel} \scrA^{-n} \\
&\propto& \scrA^{-n} \,,
\end{eqnarray}
where $B_*$ and $B_P$ are the stellar and planetary
magnetic fields, $v_{rel}$ is the relative velocity
between them that produces the shear in the magnetic
fields and leads to the reconnection.  Here, $n=3$
very close to the star and $n=2$ farther away (in the
``Parker spiral'').
Thus, if the
magnitude of the enhancement and the stellar magnetic
field can be measured for these systems, then the
planetary magnetic environment can be investigated.

Since our measurements are an ensemble average, i.e.,
they do not take into account variations in mass,
magnetic fields, and relative velocities, we cannot
directly verify the above models.  Nevertheless,
we can obtain a rough estimate for the variation of
the ``excess'' luminosity with $\scrA$ by assuming that
\begin{equation}
\delta{L_X} \propto \scrA^x
\end{equation}
where $\delta{L_X}$ is the difference between the
actual luminosity and a luminosity unaffected by
a close-in EGP.  For the former, we use the values
simulated during the Monte Carlo analysis of \S\ref{s:farout},
and approximate the latter with the average $L_X$
derived for the distant sample.  Fitting straight
lines to $\log{\delta{L_X}}$ vs $\log{\scrA}$,
and accounting for the bias in the same manner
as described above in \S\ref{s:activity}, we find
that $x\approx{-1}$.
The error on this value is however very large ($\sim 100\%$).
Therefore, we
conclude that while the data are qualitatively
consistent with the scenarios proposed by Cuntz
et al.\ (2000), a reliable test of the theory
is not feasible without more observations.


\section{Summary}
\label{s:summary}

We have searched for X-ray emission from a sample of 230
stars with known giant planets with a view to characterizing
the effect of the EGPs on the parent star.
We first find that the overall sample of stars with known
EGPs is similar in gross X-ray properties to field star
samples (see \S\ref{s:samplestat}), and thus provides a
representative sample of X-ray stars.  
We carry out a careful search for statistical trends with
various parameters (see \S\ref{s:adepend}) and find some
evidence that the activity levels of stars with close-in
giant planets is higher than for stars with planets located
further out, though the correlations are contradictory
and inconclusive.
We then carry out a more powerful test by
analyzing in detail two extremal subsamples (see \S\ref{s:farout})
where we compare the X-ray emission from nearby (distance$<$60~pc)
main-sequence stars with close-in
EGPs against similar stars with much more distant EGPs.  For the sake
of definiteness, we choose a close-in sample where the
orbital semi-major axis $\scrA<0.15$ AU, and a distant
sample where $\scrA>1.5$ AU (reducing the separation between
the samples increases the number of stars considered, but
reduces the contrast between the samples).  We verify that
our adopte ranges of $\scrA$ are not special by varying
the limiting ranges of $\scrA$ over which the subsamples
are defined, and find that invariably the close-in samples
have X-ray luminosities higher than that of the distant sample.

The above result must be understood in the context of
selection biases in our sample of stars.  In \S\ref{s:bias},
we demonstrate that observational biases account for about half
of the observed differences seen in the data.  After these
biases are accounted for, we find that the close-in
sample is more active by a factor of $2_{>1.3}^{<4}$ on
average.  This result holds even when the full data are filtered
with different selection criteria.

The robustness of the result is limited by the large number
of systems yet to be detected in X-rays.  New observations
by {\sl Chandra} or {\sl XMM-Newton} would consequently
improve the statistics and would constrain the magnitude
of this effect with better precision.  We note that a simple
model of the X-ray emission enhancement suggests an interaction
strength proportional to the product of the stellar and
planetary magnetic fields, $B_*\,B_P$ at their point of
interaction.  This suggests that if $B_*$ is known or can be
estimated, $B_P$ for exoplanets can potentially be studied.
Since the point of interaction between $B_P$ and $B_*$ depends
in part on the strength of the stellar wind, close-in wind
properties can also potentially be probed by such observations.

\acknowledgments

This research has made use of data obtained through the High Energy
Astrophysics Science Archive Research Center Online Service, provided
by the NASA/Goddard Space Flight Center, and of the SIMBAD database,
operated at CDS, Strasbourg, France.
We thank Dr.\ Jean Schneider and the Exoplanets Encyclopedia, and
the Geneva ExtraSolar Planets
group for their invaluable online compilation of EGP resources.
We also thank Pete Ratzlaff for assistance with computing counts-to-energy
conversion factors.
This work was supported by NASA grant NNG05GJ63G for XMM GO support,
NASA-AISRP grant NNG06GF17G,
and also by CXC NASA contract NAS8-39073 (VLK and JJD).
SHS was supported by NASA Origins Program Grant NAG-10360.

\newpage

\end{document}